\documentclass[10pt,journal]{IEEEtran}
\ifCLASSOPTIONcompsoc
	\usepackage[nocompress]{cite}
\else
	\usepackage{cite}
\fi
\usepackage{algpseudocode}
\usepackage{algorithm}
\usepackage[table,svgnames]{xcolor}
\usepackage{pdfpages}
\usepackage{circledsteps} 
\usepackage{setspace}
\usepackage{amsmath,cancel}
\usepackage{amssymb}
\usepackage{url}
\usepackage{mathrsfs}
\usepackage{booktabs}
\usepackage{cite}
\usepackage[table,xcdraw]{xcolor}
\usepackage{bbding}
\usepackage{upgreek}
\usepackage{arydshln}
\usepackage{colortbl}
\usepackage{caption}
\usepackage{adjustbox}
\usepackage{makecell}
\usepackage{float}
\usepackage{pifont}
\usepackage{subfig}
\usepackage{colortbl}
\usepackage[table,svgnames]{xcolor}

\newcommand{\Indent}{\hspace{2em}}  

\usepackage{
	amsfonts,
	graphicx,
	multirow,
	hhline,
	url,
	hyperref}
\newcommand\shline{\specialrule{0.8pt}{0pt}{0pt}}
\newcolumntype{I}{!{\vrule width 1pt}}

\begin{document}
	
\title{BridgeVoC: Revitalizing Neural Vocoder from a Restoration Perspective}

\author{Andong~Li, \IEEEmembership{Member,~IEEE},
		Tong Lei,
		Rilin Chen,
		Kai Li,  \IEEEmembership{Student Member,~IEEE},
		Meng Yu,
		Xiaodong Li,
		Dong~Yu, \IEEEmembership{Fellow,~IEEE},
		Chengshi~Zheng, \IEEEmembership{Senior Member,~IEEE},
		\IEEEcompsocitemizethanks{
			\IEEEcompsocthanksitem Andong Li, Xiaodong Li, and Chengshi Zheng are with the Key Laboratory of Noise and Vibration Research, Institute of Acoustics, Chinese Academy of Sciences, Beijing, 100190, China, and also with University of Chinese Academy of Sciences, Beijing, 100049, China. (Email: liandong@mail.ioa.ac.cn, lxd@mail.ioa.ac.cn, cszheng@mail.ioa.ac.cn)
			 \IEEEcompsocthanksitem Tong Lei, Rilin Chen, Meng Yu, and Dong Yu are with Tencent AI Lab.
			 \IEEEcompsocthanksitem Kai Li is with the Department of Computer Science and Technology, Institute for Artificial Intelligence, State Key Laboratory of Intelligent Technology and Systems, BNRist, IDG/McGovern Institute for Brain Research, Tsinghua Laboratory of Brain and Intelligence (THBI), Tsinghua University, Beijing, China.
			\IEEEcompsocthanksitem Corresponding author: Chengshi Zheng.
		}}
	
\markboth{Journal of \LaTeX\ Class Files,~2023}%
	{Shell \MakeLowercase{\textit{et al.}}: Bare Demo of IEEEtran.cls for Computer Society Journals}
\IEEEtitleabstractindextext{
\begin{abstract}
Despite significant advances in neural vocoders using diffusion models and their variants, these methods, unfortunately, inherently suffer from a \textit{performance-inference dilemma}, which stems from the iterative nature in the reverse inference process. This hurdle can severely impede the development of this field. To address this challenge, this paper revisits the neural vocoder task through the lens of audio restoration and propose a novel diffusion vocoder called BridgeVoC. Specifically, by rank analysis, we compare the rank characteristics of Mel-spectrum with other common acoustic degradation factors, and cast the vocoder task as a specialized case of audio restoration, where the range-space spectral (RSS) surrogate of the target spectrum acts as the degraded input. Based on that, we introduce the Schr\"odinger bridge framework for diffusion modeling, which defines the RSS and target spectrum as dual endpoints of the stochastic generation trajectory. Further, to fully utilize the hierarchical prior of subbands in the time-frequency (T-F) domain, we elaborately devise a novel subband-aware convolutional diffusion network as the data predictor, where subbands are divided following an uneven strategy, and convolutional-style attention module is employed with large kernels for efficient T-F contextual modeling. To enable single-step inference, we propose an omnidirectional distillation loss to facilitate effective information transfer from the teacher model to the student model, and the performance is improved by combining target-related and bijective consistency losses. Comprehensive experiments are conducted on various benchmarks and out-of-distribution datasets. Quantitative and qualitative results show that while enjoying fewer parameters, lower computational cost, and competitive inference speed, the proposed BridgeVoC yields state-of-the-art performance over existing advanced GAN-, DDPM- and flow-matching-based baselines with only 4 sampling steps. And consistent superiority is still achieved with single-step inference. Demo and code link: https://github.com/Andong-Li-speech/BridgeVoC-demo.
\end{abstract}
		
}

\maketitle
\IEEEdisplaynontitleabstractindextext
\IEEEpeerreviewmaketitle

\vspace{-4pt}
\section{Introduction}\label{sec:introduction}
Given acoustic features as input, audio vocoders aim to reconstruct the audible waveform by means of signal-processing and computational techniques. As a fundamental audio technology, vocoders have been widely applied to various speech and music generation and processing tasks, such as text-to-speech (TTS)~{\cite{tan2021survey,tan2024naturalspeech}}, music generation~{\cite{wu2024music,copet2023simple,bai2024seed}}, and speech enhancement~{\cite{liu2021voicefixer,shao2025cleanmel}}. Recent advances of large language models (LLMs) and generative techniques have further expanded vocoder application scenarios, including joint audio-video (AV) signal processing~{\cite{chen2024rt}}, video-to-audio generation~{\cite{li2024dance}}, and multimodal human-computer interactions~{\cite{mosqueira2023human}}.
\begin{figure}[t]
	\centering
	\vspace{0pt}
	\includegraphics[width=0.45\textwidth]{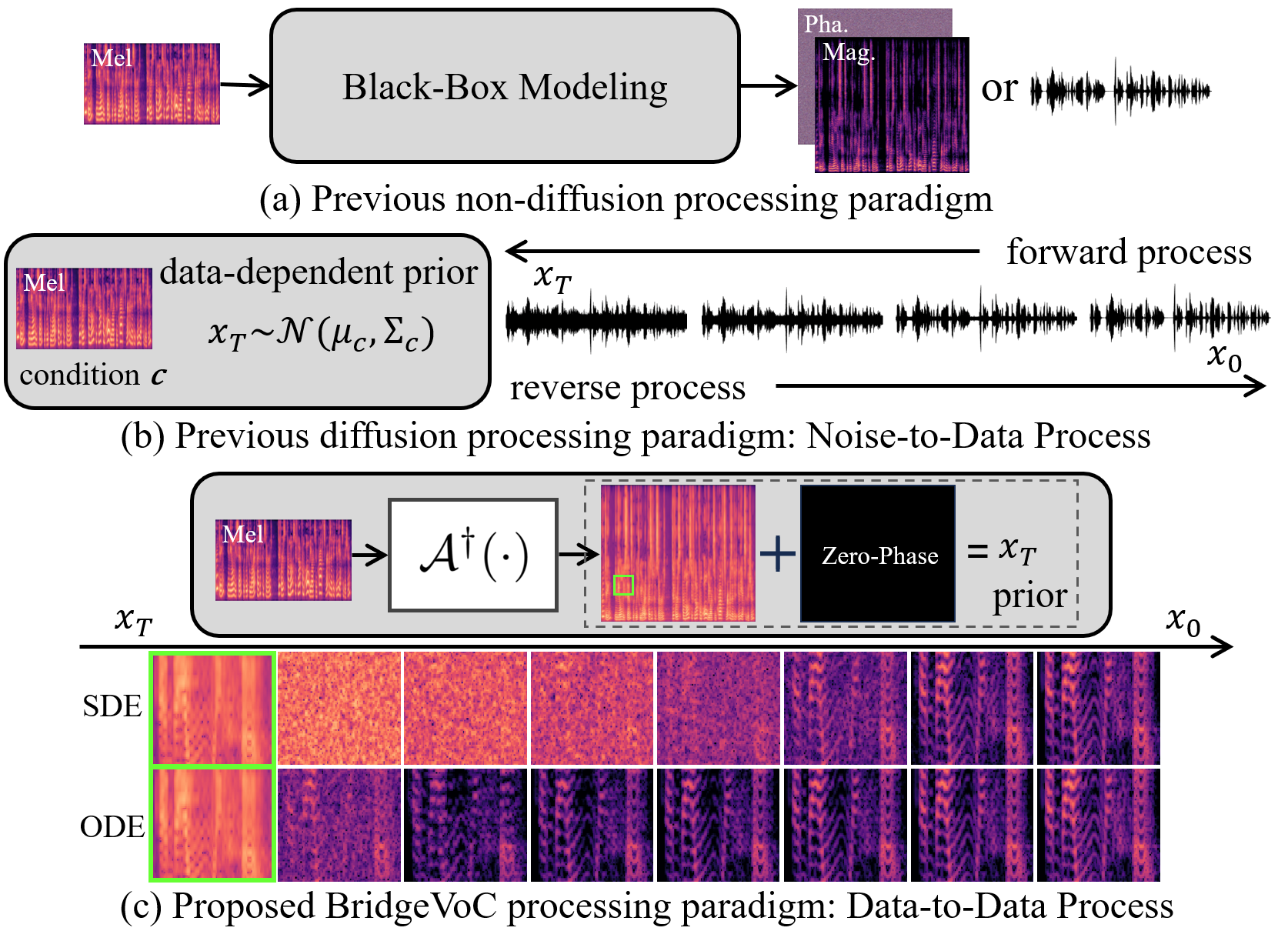}
	\vspace{0pt}
	\caption{Illustrations of different vocoder paradigms. (a) Previous non-diffusion paradigms~(\emph{e.g.}, autoregressive~{\cite{van2016wavenet}}, flow~{\cite{prenger2019waveglow}}, and GAN~{\cite{kumar2019melgan}}-based models), where the generator acts as a black-box to directly model the distribution of target waveforms in the time domain or spectrograms in the T-F domain. (b) Previous diffusion paradigms~(\emph{e.g.}, DDPM~{\cite{ho2020denoising}}, Flow Matching~{\cite{tong2023conditional}} models), where the target distribution is gradually modeled starting from Gaussian noise via a reverse process. (c) The proposed BridgeVoC, where the degradation prior of the Mel-spectrum is fully exploited, and the acoustic range-space representation serves as the source distribution.}
	\label{fig:paradigm-comparisons}
	\vspace{-0.4cm}
\end{figure}


Traditional vocoders such as STRAIGHT~{\cite{kawahara2006straight}} and WORLD~{\cite{morise2016world}}, which are based on digital signal processing (DSP), often suffer from severe distortions in spectral details and phase information, resulting in poor synthesis quality. The recent proliferation of deep neural networks has significantly enhanced the ability of deep generative models to produce high-fidelity audio waveforms, and neural vocoders have become the mainstream approach. When evaluating neural vocoders, two primary criteria should be considered:  generation quality and computational efficiency. The initial breakthroughs arose from the autoregressive generation paradigm, exemplified by WaveNet~{\cite{van2016wavenet}} and WaveRNN~{\cite{kalchbrenner2018efficient}}, which exhibit much better generation quality but incur high cost in terms of inference speed due to sample-level autoregression. To enable real-time generation, numerous non-autoregressive methods have been proposed, which can be roughly categorized into three categories: flow-based models, generative adversarial networks (GANs), and diffusion models{\footnote{Although DDPM and stochastic differential equation (SDE) are derived from different physical formulations, they both belong to the diffusion family and can be equivalently explained~\cite{luo2022understanding}. Flow Matching, by contrast, originates from normalizing continuous flows and operates by solving probability flow ordinary differential equations (ODEs). Nevertheless, recent literature has revealed their similarities and potential unification in certain scenarios~\cite{patel2024exploring,sun2025unified}. For illustration convenience, we collectively refer to them as diffusion models in this paper.}}.

\begin{figure}[t]
	\setlength{\abovecaptionskip}{-0.05cm}
	\setlength{\belowcaptionskip}{-0.05cm}
	\centering
	\vspace{0pt}
	\includegraphics[width=0.45\textwidth]{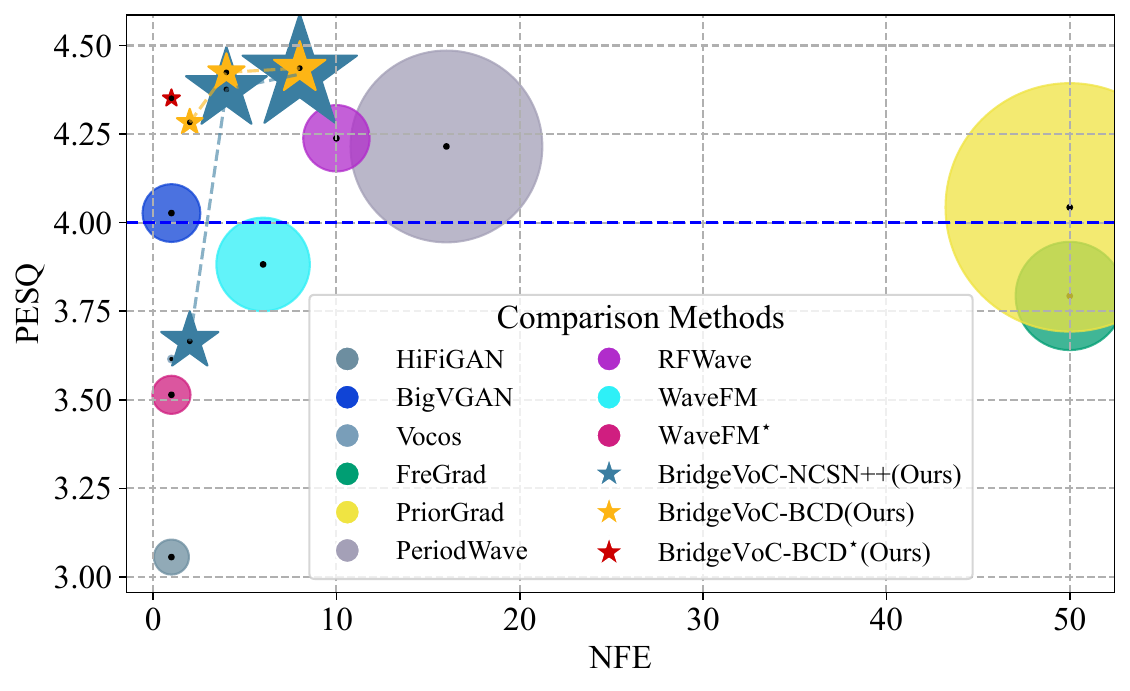}
	\caption{PESQ scores versus the number of function evaluations (NFE) for GAN-based and diffusion-based approaches on the LibriTTS benchmark. $\left(\cdot\right)^{\star}$ indicates that the single-step sampling strategy is adopted for the diffusion model. A larger bubble/star denotes higher computational complexity.}
	\label{fig:pesq-comparisons}
	\vspace{-0.4cm}
\end{figure}

Flow-based methods attempt to establish bijective mapping relationships between Gaussian and target distributions, and employ invertible neural networks that permit exact computation of the log-likelihood during training~{\cite{ping2018clarinet,prenger2019waveglow,ping2020waveflow}}. Although improved generation speed has been obtained, they incur exorbitant computational complexity during the inference stage, heavily hindering practical deployment. In contrast, GAN-based models offer greater flexibility and have demonstrated high-fidelity waveform synthesis~{\cite{kumar2019melgan,kong2020hifi,kaneko2022istftnet,siuzdakvocos}}. Conventional GAN-based vocoders typically operate in the time domain, using interleaved ResNet blocks~{\cite{he2016deep}} and upsampling operations to model target waveform distributions, while discriminators detect unnatural artifacts across multiple scales and periodicities~{\cite{kumar2019melgan,kong2020hifi}}. Recently, several studies have focused on target modeling from the time-frequency (T-F) perspective, achieving notably faster processing efficiency and competitive reconstruction performance~{\cite{siuzdakvocos,ai2023apnet,li2025neural}}. In summary, the aforementioned methods share a similar generation paradigm, as shown in Fig.~{\ref{fig:paradigm-comparisons}}(a). 

Recent breakthroughs in generative models have advanced audio generation by leaps and bounds~{\cite{ho2020denoising,rombach2022high,song2023improved}}. By constructing transformation relationships between a simple Gaussian prior and target distributions through a Markov chain process, Denoising diffusion probability models (DDPMs) achieve a high upper bound on generation quality and have been adopted for audio generation in recent years~{\cite{kongdiffwave}}. More recently, flow matching (FM)~{\cite{tong2023conditional}} has gained attention due to the simple formulation and strong performance, rendering it a viable alternative to DDPMs for neural vocoders~{\cite{lee2024periodwave,luo2025wavefm,liu2024rfwave}}. By constructing forward-backward trajectory, existing diffusion-based neural vocoders \emph{de facto} belong to the \textbf{noise-to-data} (N2D) paradigm, where the reverse process starts from a Gaussian distribution, with the Mel-spectrum serving as the conditioning input, as shown in Fig.~{\ref{fig:paradigm-comparisons}}(b).

Despite their promising performance, existing diffusion-based neural vocoders still face an inherent \textit{performance-inference dilemma}, where high quality is achieved at the cost of a relatively large number of inference steps (typically $\geq 5$). Moreover, ensuring high reconstruction quality requires large model sizes and high computational complexity. We attribute this to two primary factors:

(1) \textit{Suboptimality of the existing N2D paradigm}: Most existing studies adopt the N2D paradigm for vocoder design, \emph{i.e.}, generating audio from a simple normal distribution with the Mel-spectrum as a clue. However, it remains intrinsically challenging to reconstruct targets from pure Gaussian noise. Although strategies such as time-dependent prior variance~{\cite{leepriorgrad}} and fast-sampling~{\cite{huang2022fastdiff}} have been proposed, they yield only modest improvements. Thus, it remains questionable whether such a paradigm is an optimal choice. 

(2) \textit{Underutilization of audio prior}: While audio signals exhibit distinct spectral patterns in the T-F domain (\emph{e.g.}, harmonic components), that are well-suited for learning, existing diffusion methods typically neglect them and instead rely on complicated network structures for implicit modeling~\cite{lee2024periodwave, luo2025wavefm}, resulting in excessive resource consumption and inefficient implementation. Therefore, it remains a key challenge to incorporate audio priors for better reconstruction.

To address the first challenge, we revisit the vocoder task from a restoration perspective, and propose a \textbf{data-to-data} (D2D) paradigm for the audio generation task, as shown in Fig.~{\ref{fig:paradigm-comparisons}}(c). Specifically, we analyze the degradation characteristics of the Mel-spectrum and formulate it as a range-space spectral (RSS) surrogate of the target spectrum using range-null decomposition (RND) theory. Based on that, the original N2D process is transformed into a classical signal restoration problem, where the input is the RSS component and the target is the clean spectrum. Inspired by the success of Schr\"odinger bridge in audio restoration task, we propose a novel vocoder framework dubbed \textbf{BridgeVoC}, establishing forward-reverse transformations between the source and target distributions. By doing so, we significantly narrow the trajectory between the two endpoints. Experiments demonstrate that without adopting any fast-sampling strategies, our method achieves state-of-the-art (SoTA) performance relative to existing baselines with only 4 or even 2 inference steps, fully validating the superiority of the proposed D2D paradigm.   

To address the second challenge, inspired by the recent success of subband modeling in speech enhancement task~{\cite{li2025neural,yu2022dbt,fan2025bsdb}}, we propose a novel sub\underline{\textbf{B}}and-aware \underline{\textbf{C}}onvolutional \underline{\textbf{D}}iffusion model dubbed \textbf{BCD}, which fully leverages prior spectral knowledge. Concretely, we introduce band-division and merging strategies for hierarchical spectral encoding and decoding. Meanwhile, a convolutional-style attention block is adopted to enable efficient and effective modeling at the T-F bin level via large convolutional kernels. To further improve inference efficiency, we propose a simple yet effective strategy for one-step audio generation by incorporating target distillation and data-consistency losses. Experiments show that with a single inference step, our method achieves SoTA performance over previous baselines (see Fig.~{\ref{fig:pesq-comparisons}}). 

A precursor to this work was presented at IJCAI 2025~{\cite{tong2025bridge}}. This paper extends it with substantial advancements:\\
\noindent \ding{113}~(1) We propose a novel subband-based diffusion network called BCD, which is designed for the vocoder task. Compared with the NCSN++ architecture used in the conference version, BCD achieves better performance with significantly lower computational cost and faster inference speed.\\
\noindent \ding{113}~(2) We develop a simple yet effective scheme for single-step diffusion by integrating teacher distillation and data-consistency losses. To facilitate effective knowledge transfer from the teacher model, we propose a novel omnidirectional distillation loss that substantially improves distillation performance.\\
\noindent \ding{113}~(3) We validate the advantage of the proposed D2D paradigm on other advanced generative models, including ResShift~{\cite{yue2023resshift}} and conditional flow matching (CFM)~{\cite{tongimproving}}.

The remainder of this paper is organized as follows: Sec.~{\ref{sec:related-work}} reviews related work. Sec.~{\ref{sec:method}} presents the problem formulation and our proposed approach. Sec.~{\ref{sec:experimental-setups}} describes the datasets and experimental setup. Extensive experiments are conducted in Sec.~{\ref{sec:experiments-analysis}}. Some conclusions are drawn in Sec.~{\ref{sec:conclusion}}.
\begin{table*}[!t]
	\vspace{-5pt}
	\caption{Visualization of rank difference $\Delta\mathbf{R}$ under five types of acoustic degradations, calculated based on the VBD benchmark.}
	\vspace{-4pt}
	\large
	\label{tbl:visualize_rank}
	\centering
	\hspace{-4pt}
	\setlength{\tabcolsep}{2pt}
	\resizebox{0.84\textwidth}{!}{
		\begin{tabular}{l|ccccc}
			\shline
			\makebox[2.2cm][c]{Degrade.} &Environmental &Room &\multirow{2}{*}{Low-pass Filter} &\multirow{2}{*}{Audio Codec} 
			&\multirow{2}{*}{\textbf{Vocoder}} \\
			\makebox[2cm][c]{Type} &Noise &Reverberation & & &\\ \hline
			\makebox[2cm][c]{\multirow{4}*{Model}} 		&$\mathbf{y} = \mathbf{x} + \mathbf{n}$ 
			&$\mathbf{y} = \mathbf{x}\otimes\mathbf{h}$ 
			&$\mathbf{y} = \mathcal{H}\left(\mathbf{x}\right)$ 
			&$\mathbf{y} = \mathcal{D}\left(\mathcal{Q}\left(\mathcal{E}\left(\mathbf{x}\right)\right)\right)$
			&$\mathbf{Y} = \mathcal{A}^{\dagger}\mathbf{Z}\exp\left(\mathbf{0}\right)$\\ 
			&\multirow{3}*{$\mathbf{n}$: Noise signals} &\multirow{1}*{$\mathbf{h}$: Room impulse response} &\multirow{3}*{$\mathcal{H}\left(\cdot\right)$: Low-pass filter} &$\mathcal{E}\left(\cdot\right)$: Encoder &$\mathcal{A}^{\dagger}$: pseudo-inverse of $\mathcal{A}$ \\
			& &\multirow{2}*{$\otimes$: Convolution operation} & &$\mathcal{Q}\left(\cdot\right)$: Quantizer &\multirow{2}*{$\mathbf{Z}$: Mel-spectrum} \\
			& & & &$\mathcal{D}\left(\cdot\right)$: Decoder & \\
			\hline
			\makebox[2cm][c]{\rotatebox{90}{\centering  Visualization}} &\includegraphics[width=0.25\textwidth]{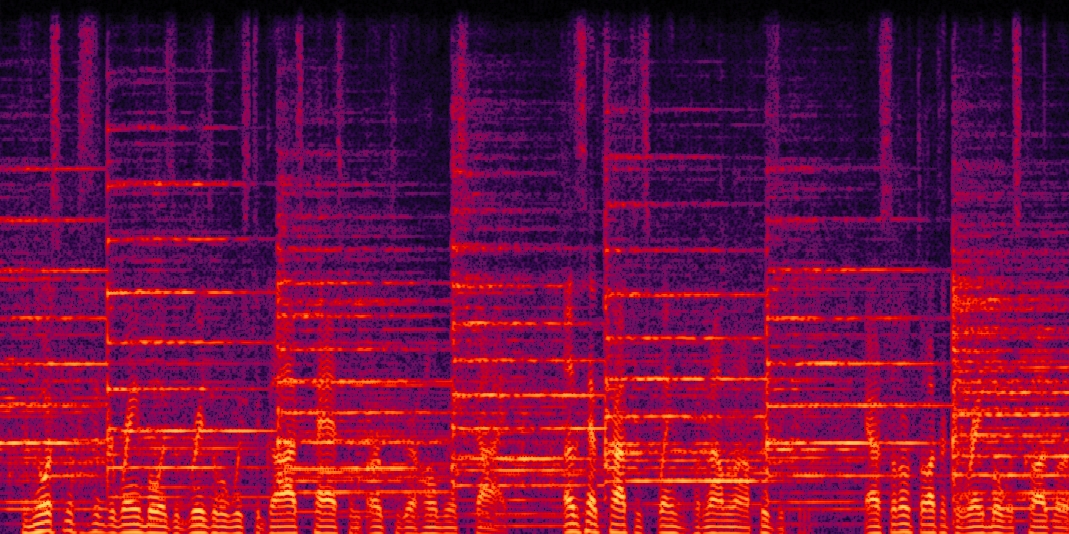} 
			&\includegraphics[width=0.25\textwidth]{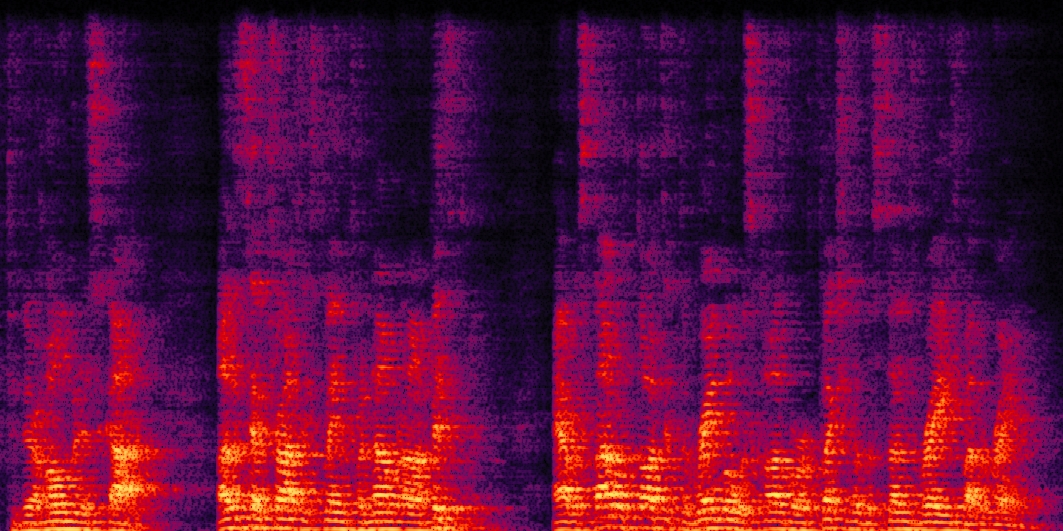} 
			&\includegraphics[width=0.25\textwidth]{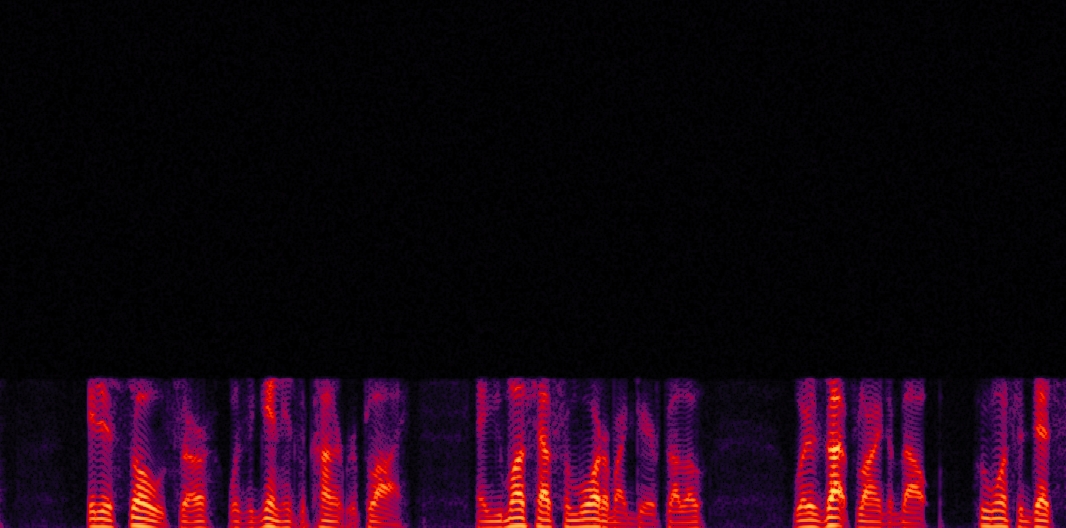}
			&\includegraphics[width=0.25\textwidth]{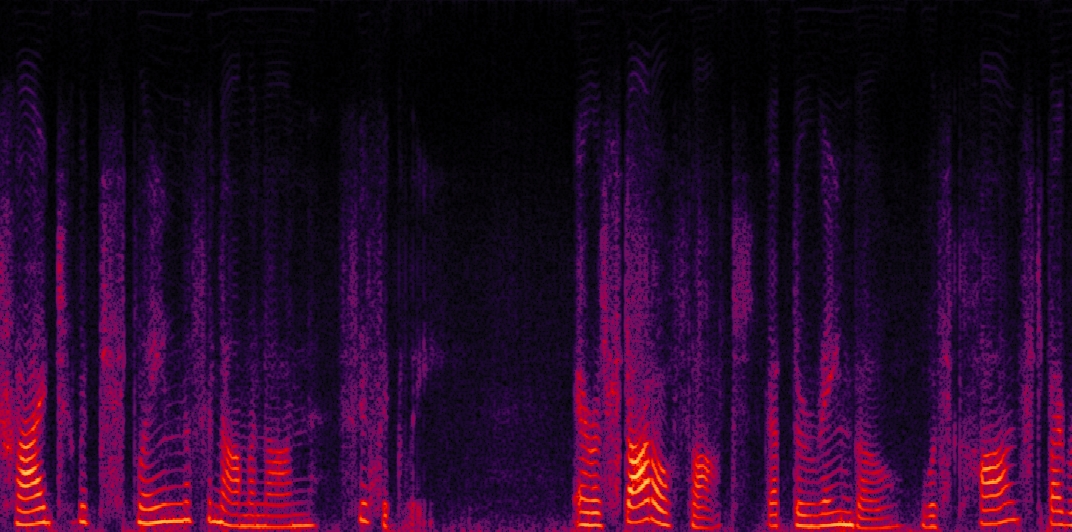}
			&\includegraphics[width=0.25\textwidth]{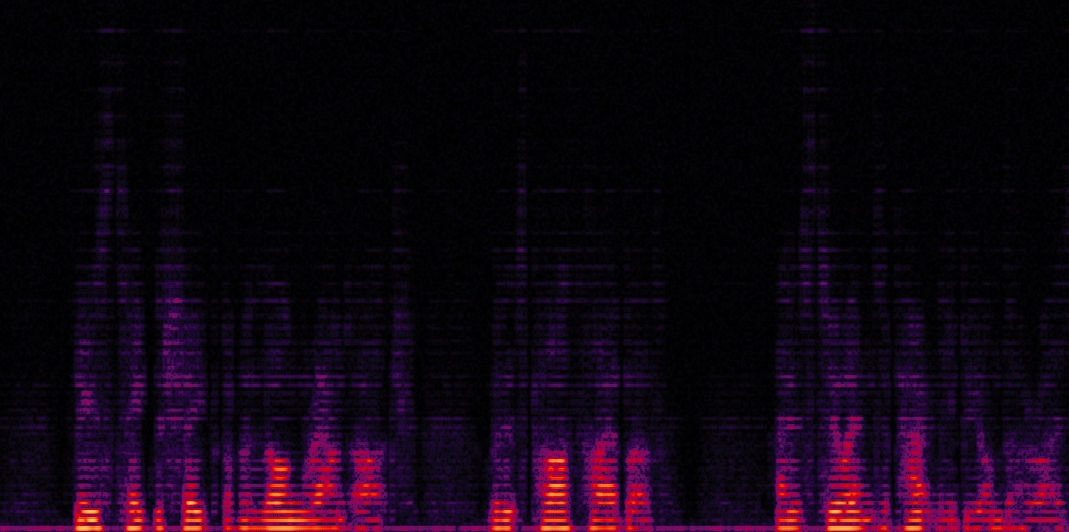}\\ \hline
			\makebox[2cm][c]{\rotatebox{90}{\centering Rank Difference}} &\includegraphics[width=0.289\textwidth]{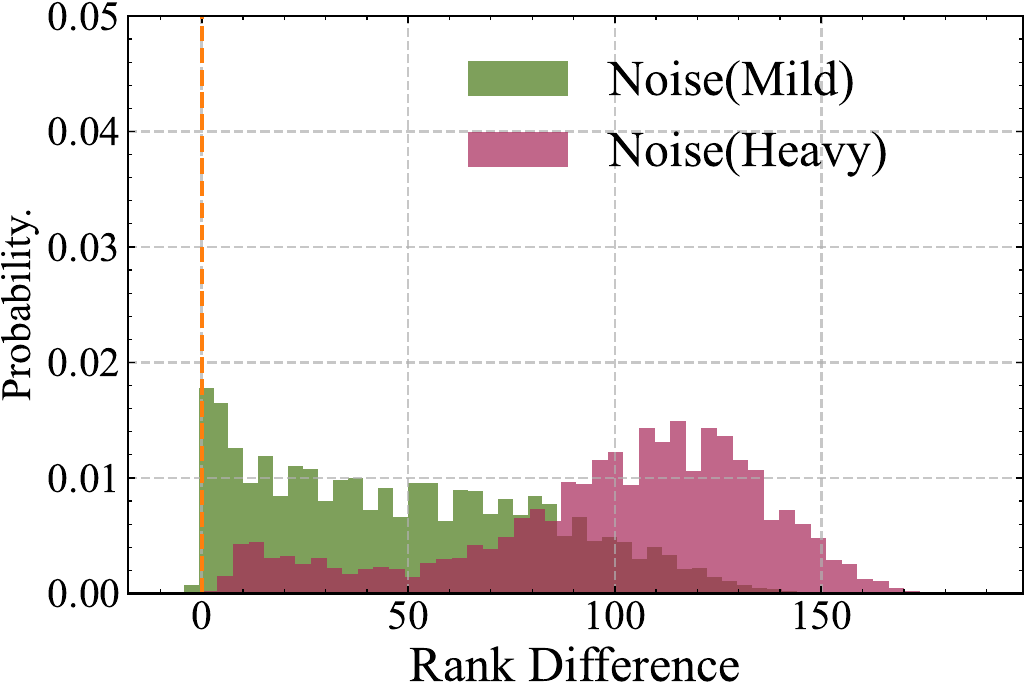} 
			&\includegraphics[width=0.26\textwidth]{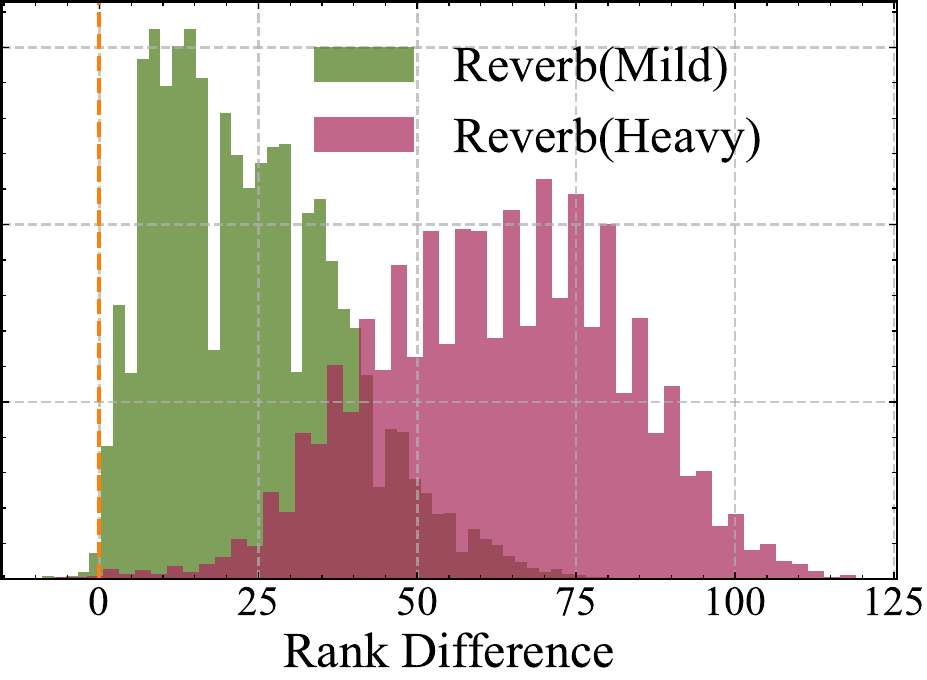} 
			&\includegraphics[width=0.25\textwidth]{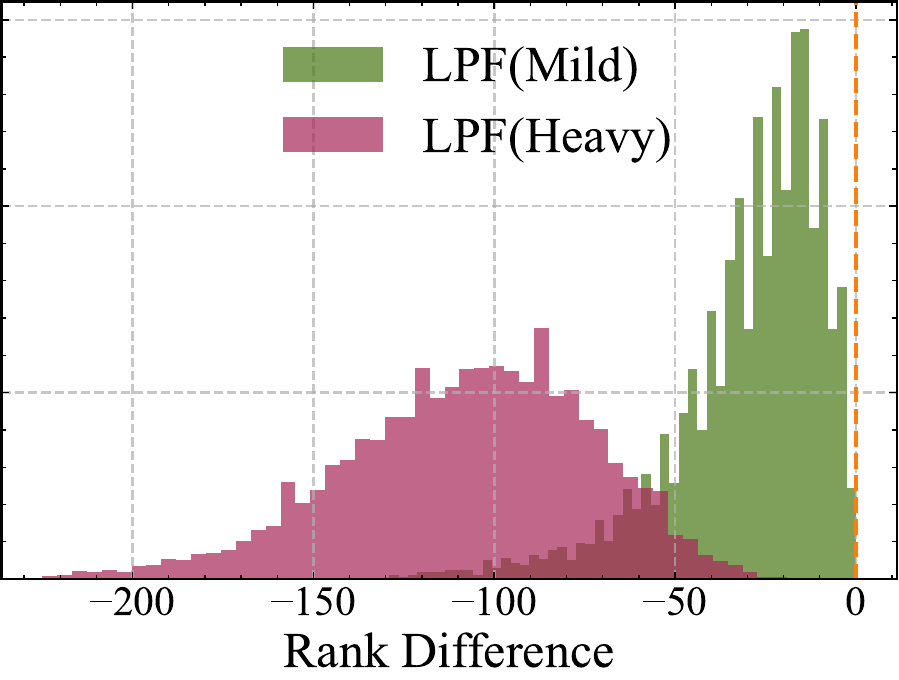}
			&\includegraphics[width=0.25\textwidth]{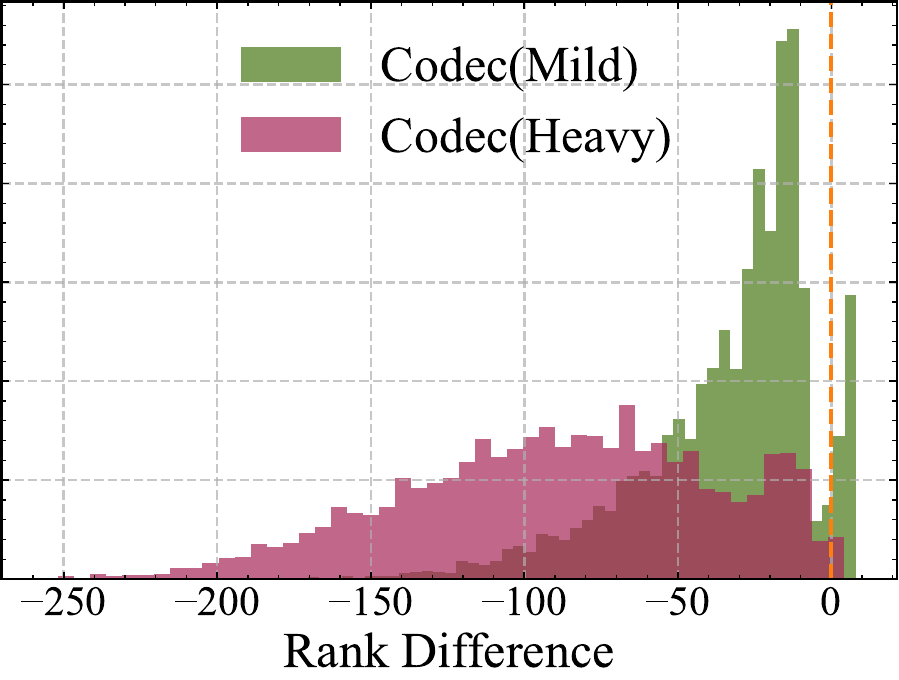}
			&\includegraphics[width=0.25\textwidth]{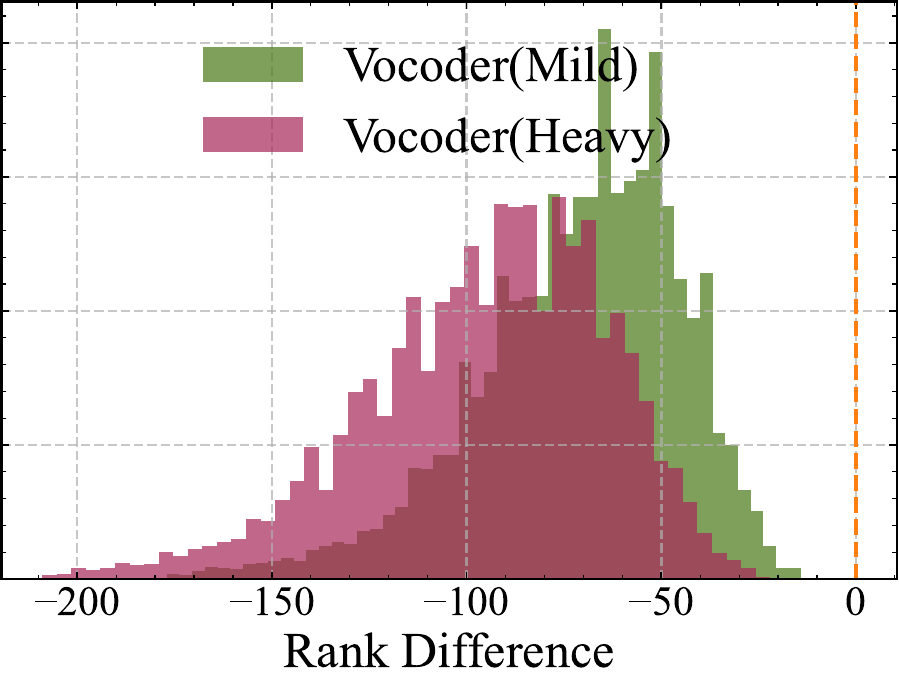}\\
			\shline
		\end{tabular}	}
	\vspace{-4pt}
\end{table*}
\vspace{-8pt}
\section{Related Work}
\label{sec:related-work}
\subsection{Non-diffusion-based Vocoder Methods}\label{sec:non-diffusion-methods}
\noindent\textit{Traditional Methods}: Traditional DSP-based vocoders typically employ an analysis-synthesis pipeline that represents audio using statistical parameters. In~{\cite{kawahara2006straight}}, excitation and resonant features are first extracted separately, followed by waveform manipulation and synthesis. Morise \emph{et al.}~{\cite{morise2016world}} proposed a real-time vocoder that first estimates the fundamental frequency (F$_{0}$), spectral envelope and aperiodic parameters via analysis algorithms, then uses a synthesis algorithm to generate audio.\\
\noindent\textit{Autoregressive-based Methods}: Autoregressive vocoders generate audio at the sample level. In~{\cite{van2016wavenet}}, WaveNet was introduced to predict each sample based on previous samples. Valin \emph{et al.}~{\cite{valin2019lpcnet}} proposed LPCNet, which estimates samples under the guidance of linear predictive coding (LPC) coefficients. Despite improvements in audio quality, these methods suffer from slow generation speeds, limiting their deployment in practical scenarios.\\
\noindent\textit{Flow-based Methods}: Flow-based vocoders aim to establish a bijective mapping between a Gaussian and target distributions. Invertible network architectures are devised to enable direct optimization of the log-likelihood function. Representative works include WaveGlow~{\cite{prenger2019waveglow}}, and FlowWaveNet~{\cite{kim2019flowavenet}}. Although these methods improve generation speed compared to autoregressive models, they face performance limitations due to constraints in network design. Furthermore, high inference cost is often required, resulting in low generation efficiency.\\
\noindent\textit{GAN-based Methods}: GAN-based vocoders offer greater flexibility in network design and training strategies, thanks to adversarial training. Existing literature can be usually delineated into two categories: time-domain-based and T-F-domain based models. For the former, Kong \emph{et al.}~{\cite{kong2020hifi}} devised two types of discriminators to ensure accurate waveform reconstruction across various sample scales and periods. Lee \emph{et al.}~{\cite{leebigvgan}} introduced a snake activation function to enhance the generation of periodic components, and scaled the model size to 112 M parameters, achieving impressive performance. For the latter, Fourier priors are incorporated to reduce modeling complexity. Siuzdak~{\cite{siuzdakvocos}} adopted ConvNext blocks for simultaneous estimation of spectral magnitude and phase. Similarly, a parallel network structure was proposed for joint magnitude and phase estimation~{\cite{ai2023apnet}}.
\vspace{-12pt}
\subsection{Diffusion-based Vocoder Methods}
\label{sec:diffusion-based-vocoder-methods}
With the rapid development of diffusion models, they have been increasingly applied to vocoder tasks. Kong \emph{et al.}~{\cite{kongdiffwave}} pioneered this direction by using DDPMs to gradually model the target waveform distribution. Later, to better guide the generation trajectory, Lee \emph{et al.}~{\cite{leepriorgrad}} incorporated frame-level Mel-spectrum energy as a prior to modify the standard Gaussian distribution. In~{\cite{nguyen2024fregrad}}, a lightweight diffusion vocoder was proposed, which incorporates discrete wavelet transform and several optimization tricks for improved performance. 

Recently, Flow Matching (FM) has gained increasing attention due to its simple formulation and stable performance. Lee \emph{et al.}~{\cite{lee2024periodwave}} introduced Optimal Transport FM (OTFM) and incorporated a period-aware estimation strategy to generate high-quality waveforms. Luo \emph{et al.}~{\cite{luo2025wavefm}} proposed Rectified FM (RFM) and used consistency distillation~{\cite{song2023consistency}} to enable single-step generation, though its performance remained modest. Unlike the aforementioned time-domain diffusion methods, Liu \emph{et al.}~{\cite{liu2024rfwave}} proposed a T-F-domain diffusion vocoder based on RFM. This model shares a ConvNeXt-based velocity field estimator across multiple subbands, achieving a notable inference speed advantage over~{\cite{leepriorgrad,nguyen2024fregrad}} when using 10 inference steps. 
\vspace{-0.3cm}
\section{Method}\label{sec:method}
\subsection{Problem Formulation}\label{sec:problem-formulation}
\subsubsection{Range-Null Decomposition}\label{sec:rnd}
We begin with a classical signal compression model:
\begin{equation}
	\label{eqn:1}
	\setlength{\abovedisplayskip}{3pt}
	\setlength{\belowdisplayskip}{4pt}
	\mathbf{y} = \mathbf{A}\mathbf{x} + \mathbf{n},
\end{equation}
where $\left\{\mathbf{x}\in\mathbb{R}^{D}, \mathbf{n}\in\mathbb{R}^{d}, \mathbf{y}\in\mathbb{R}^{d}\right\}$ denotes the target, noise, and observed signals, respectively. $\mathbf{A}\in\mathbb{R}^{d\times D}$ is a linear compression matrix. The feature dimensions $\left\{D, d\right\}$ typically satisfy $d\ll D$. For the noise-free scenario, \emph{i.e.}, $\mathbf{y} = \mathbf{A}\mathbf{x}$, we define $\mathbf{A}^{\dagger}\in\mathbb{R}^{D\times d}$ as the pseudo-inverse form of $\mathbf{A}$, which satisfies $\mathbf{A}\mathbf{A}^{\dagger}\mathbf{A}\equiv\mathbf{A}$. Using range-null decomposition (RND), $\mathbf{x}$ can be decomposed into:
\begin{equation}
	\label{eqn:2}
	\setlength{\abovedisplayskip}{3pt}
	\setlength{\belowdisplayskip}{4pt}
	\mathbf{x} \equiv \mathbf{A}^{\dagger}\mathbf{A}\mathbf{x} + \left(\mathbf{I} - \mathbf{A}^{\dagger}\mathbf{A}\right)\mathbf{x},
	\vspace{-3pt}
\end{equation}
where $\mathbf{A}^{\dagger}\mathbf{A}\mathbf{x}$ denotes the range-space (RS) component, and $\left(\mathbf{I} - \mathbf{A}^{\dagger}\mathbf{A}\right)\mathbf{x}$ is the null-space (NS) component. These two components are orthogonal. For the compression model above, we require the solution $\tilde{\mathbf{x}}$ to satisfy two criteria: (1) \textit{Degradation consistency}, which ensures the estimation to preserve all effective priors of $\mathbf{y}$; (2) \textit{Data consistency}, \emph{i.e.}, $\tilde{\mathbf{x}}\sim p\left(\mathbf{x}\right)$, which guarantees generation quality. With these two constraints, the solution $\tilde{\mathbf{x}}$ can be expressed as:
\begin{equation}
	\label{eqn:3}
	\setlength{\abovedisplayskip}{3pt}
	\setlength{\belowdisplayskip}{4pt}
	\tilde{\mathbf{x}} = \mathbf{A}^{\dagger}\mathbf{y} + \left(\mathbf{I} - \mathbf{A}^{\dagger}\mathbf{A}\right)\mathbf{\hat{x}}.
	\vspace{-3pt}
\end{equation}

Notably, this solution satisfies the first constraint due to orthogonality, as shown by,
\begin{equation}
	\label{eqn:4}
	\setlength{\abovedisplayskip}{3pt}
	\setlength{\belowdisplayskip}{4pt}
	 \mathbf{A}\tilde{\mathbf{x}}=\mathbf{A}\mathbf{A}^{\dagger}\mathbf{A}\mathbf{x} + \mathbf{A}\left(\mathbf{I} - \mathbf{A}^{\dagger}\mathbf{A}\right)\mathbf{\hat{x}} = \mathbf{A}\mathbf{x},
	 \vspace{-3pt}
\end{equation}

To satisfy the second condition, we attempt to model the NS component to approximate that of the target signal, which can be achieved using various generative methods.  
\subsubsection{Signal Models}\label{sec:signal-models}
The physical signal model for the Mel-spectrum is given by:
\begin{equation}
	\label{eqn:5}
	\setlength{\abovedisplayskip}{3pt}
	\setlength{\belowdisplayskip}{4pt}
	\mathbf{X}^{mel} = \log\left(\mathcal{A}\left|\mathbf{X}\right|\right),
	\vspace{-3pt}
\end{equation} 
where $\mathbf{X}\in\mathbb{C}^{F\times L}$ denotes the target spectrum in the T-F domain, with $\left\{F, L\right\}$ representing the frequency and frame sizes, respectively. $\left|\cdot\right|$ is the modulus operation. $\mathcal{A}\in\mathbb{R}^{F_{m}\times F}$ is the Mel-filter bank, which is instantiated as a linear compression matrix. $\mathbf{X}^{mel}\in\mathbb{R}^{F_{m}\times L}$ is the resultant log-scale Mel-spectrum, where $F_{m}$ is the frequency size of Mel-spectrum. Absorbing the logarithm operation into the left-hand side of Eq.~{\ref{eqn:5}} yields:
\begin{equation}
	\label{eqn:6}
	\setlength{\abovedisplayskip}{3pt}
	\setlength{\belowdisplayskip}{4pt}
	\mathbf{Z} = \exp\left(\mathbf{X}^{mel}\right) = \mathcal{A}\left|\mathbf{X}\right|.
	\vspace{-3pt}
\end{equation} 

Compared to the target spectrum $\mathbf{X}$, the Mel-spectrum $\mathbf{Z}$ usually suffers from two key degradations: $\Circled{\footnotesize{1}}$ Loss of phase information $\mathbf{\Phi}$; $\Circled{\footnotesize{2}}$  Magnitude information loss via linear matrix compression. Correspondingly, waveform generation from the Mel-spectrum can be transformed into a signal inverse problem involving two sub-tasks:  
\begin{itemize}
	\item \textit{Phase retrieval}: A classical signal processing problem widely studied in audio~{\cite{peer2022beyond}} and optics~{\cite{stolle1996phase}}. Recent works have proposed neural networks (NN)-based methods for improved prediction accuracy~{\cite{peer2023diffphase,dai2025bapen}}. 
	\item \textit{Magnitude recovery}: Involving reconstructing the target from linearly compressed observations, with applications in compressive sensing~{\cite{baraniuk2010model}} and computational imaging systems~{\cite{huang2023deep}}.
\end{itemize}

Notably, Eq.~{\ref{eqn:6}} shares a similar form to the noise-free case of Eq.~{\ref{eqn:1}}, motivating us to bridge the connection between the vocoder task with RND. Following Eq.~{\ref{eqn:3}}, the estimation of the target spectrum can be written as:
\begin{equation}
	\label{eqn:7}
	\setlength{\abovedisplayskip}{3pt}
	\setlength{\belowdisplayskip}{4pt}
	\left|\tilde{\mathbf{X}}\right| = \mathcal{A}^{\dagger}\mathbf{Z} + \left(\mathbf{I} - \mathcal{A}^{\dagger}\mathcal{A}\right) \left|\hat{\mathbf{X}}\right|,
	\vspace{-3pt}
\end{equation}
where $\mathcal{A}^{\dagger}\in\mathbb{R}^{F\times F_{m}}$ denotes the pseudo-inverse of $\mathcal{A}$. In this way, the magnitude recovery sub-task is transformed into a restoration problem from the RS representation, \emph{i.e.}, $\mathcal{A}^{\dagger}\mathbf{Z}\rightarrow\left|\mathbf{X}\right|$. Note that Eq.~{\ref{eqn:7}} does not involve phase information. To address this, we empirically initialize the phase as all-zero tensor $\mathbf{\Phi}_{init} = \mathbf{0}$, leading to the following formulation of the signal inverse problem:
\begin{equation}
	\label{eqn:8}
	\setlength{\abovedisplayskip}{3pt}
	\setlength{\belowdisplayskip}{4pt}
	\left\{\left|\tilde{\mathbf{X}}\right|, \tilde{\mathbf{\Phi}}\right\} = \mathcal{F}\left(\mathcal{A}^{\dagger}\mathbf{Z}\exp\left(\mathbf{\Phi}_{init}\right)\right),
	\vspace{-3pt}
\end{equation}
where $\mathcal{F}\left(\cdot\right)$ denotes the processing function that takes the initialized input $\mathcal{A}^{\dagger}\mathbf{Z}\exp\left(\mathbf{\Phi}_{init}\right)$ and outputs the estimated magnitude and phase.
\begin{table}
	\centering
	\caption{The noise schedules used in the Schr\"odinger Bridge. $\Delta\beta = \beta_{1} - \beta_{0}$, and $\left\{\beta_{0}, \beta_{1}\right\}$ are empirically set to $\left\{0.01, 20\right\}$ for the gmax schedule; For the variance-preservation (VP) schedule, $\left\{\beta_{0}, c\right\} = \left\{0.01, 0.4\right\}$; For the variance-exploding schedule, $\left\{c, k\right\} = \left\{0.01, 2.6\right\}$. The choices of these parameters follow the previous literature~{\cite{chen2023schrodinger}} and also perform well in our internal experiments.}
	\resizebox{0.49\textwidth}{!}{
		\begin{tabular}{c|ccc}
			\hline
			\multicolumn{1}{c}{Scheduler} & \multicolumn{1}{c}{gmax} & \multicolumn{1}{c}{VP} & \multicolumn{1}{c}{VE} \\
			\hline
			\(f(t)\)      & 0            & \(-\frac{1}{2}(\beta_0+t\Delta\beta)\)& 0  \\
			\(g^2(t)\)    & \(\beta_0+t\Delta\beta\)& \(c(\beta_0+t\Delta\beta)\)&\(ck^{2t}\) \\
			\(\alpha_t\)  & 1            & \(e^{-\frac{1}{2}\int_0^t(\beta_0+\tau\Delta\beta)\mathrm{d}\tau}\)& 1  \\
			\(\sigma_t^2\)& \(\frac{t^2\Delta\beta}{2}+\beta_0t\)& \(c(e^{\int_0^t(\beta_0+\tau\Delta\beta)\mathrm{d}\tau}-1)\)& \(\frac{c\left(k^{2t}-1\right)}{2\log(k)}\) \\
			\hline
	\end{tabular}}
	\label{tbl:noiseschedules}
	\vspace{-12pt}
\end{table}

\setcounter{figure}{3}
\begin{figure*}
	\centering
	\vspace{0pt}
	\includegraphics[width=0.8\textwidth]{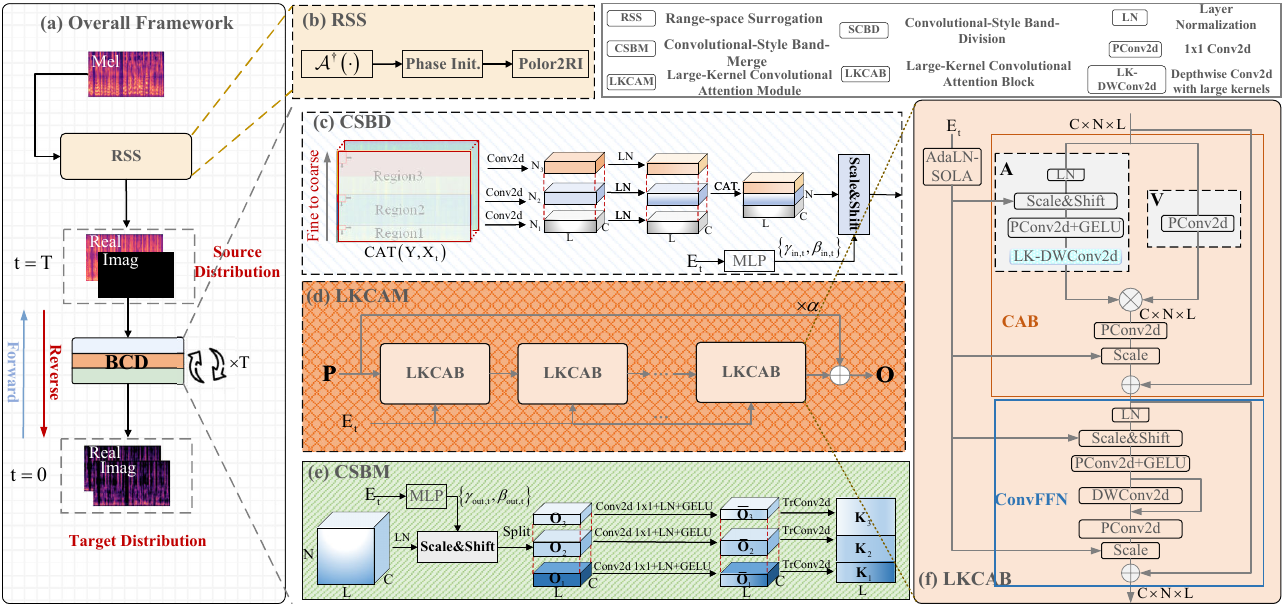}
	\vspace{-2pt}
	\caption{Framework of the proposed BridgeVoC, where BCD serves as the data predictor. (a) Overall forward and reverse process architecture. (b) Details of range-space spectral surrogate. (c) Internal structure of the proposed convolutional-style subband-division module (CSBD). (d) Internal structure of the proposed large-kernel convolutional attention module (LKCAM), compoased of stacked LKCABs. (e) Internal structure of the proposed convolutional-style band-merge module (CSBM). (f) Internal structure of the adopted LKCAB, which includes a convolutional attention block (CAB) and a convolutional feed-forward network (ConvFFN). A conditional time embedding $\mathbf{E}_{t}$ is introduced for feature modulation at each iterative step.}
	\label{fig:framework}
	\vspace{-0.45cm}
\end{figure*}
\subsubsection{Rank Analysis}\label{sec:rank-analysis}
In Sec.~{\ref{sec:signal-models}}, we modeled the vocoder task as a signal inverse problem. Here, we further analyze its rank behavior and compare it with other speech processing tasks. Recall that speech enhancement has been extensively studied over the past decades~{\cite{zheng2023sixty}}, which involves the enhancement process from various acoustic degradations, \emph{e.g.}, environmental noise, room reverberation and low-pass filter. To reveal the connection between vocoding and restoration tasks, we introduce \textbf{rank analysis} as a metric. Specifically, given the magnitude spectra of the degraded and target signals, \emph{i.e.}, $\left\{\left|\mathbf{Y}\right|, \left|\mathbf{X}\right|\right\}$, the rank difference (RD) can be defined as:
\begin{equation}
	\label{eqn:9}
	\Delta{\mathbf{R}} = \mathcal{R}\left(\left|\mathbf{Y}\right|\right) - \mathcal{R}\left(\left|\mathbf{X}\right|\right),
\end{equation}
where $\mathcal{R}\left(\cdot\right)$ denotes the matrix rank operation{\footnote{Phase is excluded from rank calculation here, as the majority of information energy resides in the magnitude spectrum.}}. In Table~{\ref{tbl:visualize_rank}}, we present histograms of $\Delta\mathbf{R}$ for four common degradations as well as the vocoder task formulated in Sec.~{\ref{sec:signal-models}}. The VoiceBank+DEMAND (VBD)~{\cite{veaux2013voice}} benchmark with 16 kHz sampling rate is utilized. For each degradation type, we consider two severity levels, namely mild and heavy, and details are listed below:
\begin{itemize}
	\item Noise: Noises are from the noise set in the VBD benchmark. The mild version follows the original VBD signal-to-noise (SNR) configurations. For the heavy case, we reduce the overall SNR by 10 dB.
	\item Reverberation: Room impulse responses (RIRs) are generated using the gpuRIR package~{\cite{diaz2021gpurir}}. For mild version, the reverberation time (RT$_{60}$)$<$0.6 s, and RT$_{60}$ $\ge$0.6 s for heavy case. 
	\item Low-pass filter (LPF): For the mild case, the cutoff frequency is set to 4 kHz, and 1 kHz for heavy case.
	\item Audio Codec: The Opus audio codec{\footnote{https://github.com/xiph/opus}} is adopted. For the mild version, the bitrate is set to around 6.4 kbps, and 2 kbps for heavy case.
	\item Vocoder: The Mel size $F_{m}$ is set to 80 for mild version, and 64 for heavy version.
\end{itemize}

Table~{\ref{tbl:visualize_rank}} reveals several key observations. First, \textit{noise and room reverberation can increase spectral rank}, \emph{i.e.}, $\Delta\mathbf{R}\ge 0$, which can be explained as additional interference is introduced, \emph{e.g.}, noise or early/late reverberation components. As a result, the information filtering mechanism is usually required for the reverse process. Second, \textit{LPF and audio codec can reduce the spectral rank}, $\Delta\mathbf{R}\le 0$, as high-frequency components are usually distorted or removed, leading to more sparseness in the spectrum. Thus, an information generation mechanism is required for the reverse process. Third, \textit{the Mel-formulation exhibits rank behavior similar to LPF and codec}, which can be formally proven as:
\begin{equation}
	\label{eqn:10}
	\setlength{\abovedisplayskip}{3pt}
	\setlength{\belowdisplayskip}{4pt}
	\mathcal{R}\left(\left|\mathbf{Y}\right|\right) = \mathcal{R}\left(\mathcal{A}^{\dagger}\mathcal{A}\left|\mathbf{X}\right|\right)\leq\min\left\{\mathcal{R}\left(\mathcal{A}^{\dagger}\mathcal{A}\right), \mathcal{R}\left(\left|\mathbf{X}\right|\right)\right\}.
	\vspace{-3pt}
\end{equation}
\setcounter{figure}{2}
\begin{figure}[t]
	\centering
	\vspace{0pt}
	\includegraphics[width=0.40\textwidth]{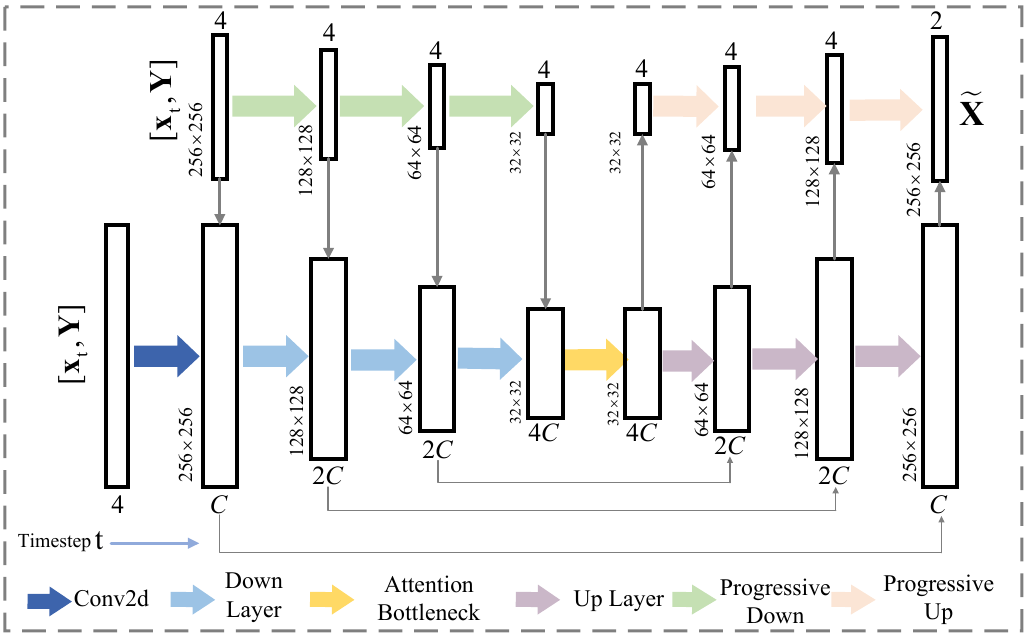}
	\vspace{-2pt}
	\caption{Overall structure of the previously adopted NCSN++ model. $C$ denotes the original number of feature channels.}
	\label{fig:ncsn_system}
	\vspace{-0.6cm}
\end{figure}

Recently, universal speech enhancement (USE)~{\cite{serra2022universal}} has gained significant attention for its remarkable capability to address diverse signal degradations within a unified framework. From this perspective, we argue that \textbf{the vocoder task can be regarded as a specialized audio restoration task; accordingly, methodologies and design principles from USE, \emph{e.g.}, network structure or processing paradigms, can be adapted}.
\vspace{-0.5cm}
\subsection{Score-based Generative Models}\label{sec:score-based-diffusion}
\vspace{-4pt}
Recently, score-based generative models (SGMs)~{\cite{song2020score}} have been applied to different SE tasks, which rely on a continuous-time diffusion process defined by a forward stochastic differential equation (SDE):
\begin{align}
	\label{eqn:11}
	\setlength{\abovedisplayskip}{3pt}
	\setlength{\belowdisplayskip}{4pt}
	\mathrm{d}\boldsymbol{x}_t=\boldsymbol{f}(\boldsymbol{x}_t,t)\mathrm{d}t+g(t)\mathrm{d}\boldsymbol{w}_t,\quad\boldsymbol{x}_0\sim p_0=p_\mathrm{data},
\end{align}
where $p_\mathrm{data}$ denotes the data distribution, $\boldsymbol{x}_{t}\in\mathbb{R}^{D}$ is the state at time $t$, and $t\in[0, T]$ is a finite time index. $\boldsymbol{f}$ is a vector-valued drift term, $g$ is a scalar-valued diffusion term, and $\boldsymbol{w}_{t}\in\mathbb{R}^{D}$ is a standard Wiener process. For the case when the boundary distribution is a prior zero-mean Gaussian $p_{\text{prior}}\sim\mathcal{N}\left(0, \sigma_{T}^{2}\boldsymbol{I}\right)$, the reverse SDE can be written as:
\begin{align}
	\label{eqn:12}
	\setlength{\abovedisplayskip}{3pt}
	\setlength{\belowdisplayskip}{4pt}
	\mathrm{d}\boldsymbol{x}_t=[\boldsymbol{f}(\boldsymbol{x}_t,t)-g^2(t)\nabla\log p_t(\boldsymbol{x}_t)]\mathrm{d}t+g(t)\mathrm{d}\bar{\boldsymbol{w}}_t,\nonumber \\
	\boldsymbol{x}_T\sim p_T\approx p_\mathrm{prior},
\end{align}
where $\bar{\boldsymbol{w}}_t$ is a reverse-time Wiener process, and $\nabla\log p_t(\boldsymbol{x}_t)$ is the score function of the marginal distribution $p_{t}$. To solve Eq.~{\ref{eqn:12}} iteratively, starting from $p_\mathrm{prior}$ at $t=T$, we replace the intractable $\nabla\log p_t(\boldsymbol{x}_t)$ with a trainable score network $s_\theta(\boldsymbol{x}_t, t)$, and a denoising score matching (DSM) training objective is given by:
\begin{equation}
	\label{eqn:13}
	\setlength{\abovedisplayskip}{3pt}
	\setlength{\belowdisplayskip}{4pt}
	\mathbb{E}_{p_0(\boldsymbol{x}_0)p_{t|0}(\boldsymbol{x}_t|\boldsymbol{x}_0), t}\left[\lambda_{t}\|s_\theta(\boldsymbol{x}_t, t) - \nabla \log p_{t|0}(\boldsymbol{x}_t|\boldsymbol{x}_0)\|_2^2\right],
	\vspace{-3pt}
\end{equation}
where $\mathbb{E}\left[\cdot\right]$ denotes the mathematical expectation, $t\sim\mathcal{U}(0,T)$, and \(p_{t|0}\) is the conditional transition distribution from \(\boldsymbol{x}_0\) to \(\boldsymbol{x}_t\), determined by the pre-defined forward SDE and analytical for a linear drift \(\boldsymbol{f}(\boldsymbol{x}_t,t) = f(t)\boldsymbol{x}_t\). $\lambda_{t}$ is a weighting coefficient, empirically set to 1. 
\vspace{-12pt}
\subsection{Schr\"odinger Bridge}\label{sec:schrodinger-bridge}
\vspace{-4pt}
Instead of starting from a standard Gaussian distribution in conventional diffusion models~{\cite{ho2020denoising,song2020score}}, Schr\"odinger Bridge (ScB)~{\cite{chen2021likelihood}} effectively models dual distributions between a source and target, which has been widely applied to image translation~{\cite{kim2024latent}} and speech synthesis ~{\cite{chen2023schrodinger}}. Recall from Sec.~{\ref{sec:problem-formulation}} that we derived a decent surrogate for the Mel-spectrum, \emph{i.e.}, the RSS representation $\mathbf{Y} = \mathcal{A}^{\dagger}\mathbf{Z}\exp\left(\mathbf{\Phi}_{init}\right)$. From a restoration perspective, ScB is therefore well-suited to model the forward-reverse process between $\mathbf{Y}$ and $\mathbf{X}${\footnote{We emphasize that the ScB method is not the only optimal choice for source-target distribution modeling. However, it performs effectively in our experiments and achieves SoTA results. Besides, in Sec.~{\ref{sec:performance-different-diffusion}}, we extend our framework to other diffusion models like ResShift and CFM, and observe comparable performance, further validating the feasibility of the proposed restoration perspective for the vocoder task.}}.

The ScB problem optimizes path measures with constrained boundary distributions. For the vocoder task, we define the target distribution $p_{\mathbf{X}}$ as the data distribution $p_{\text{data}}$ and the source distribution $p_{\mathbf{Y}}$ as the prior distribution. Considering $\left\{p_{0}, p_{T}\right\}$ as the marginal distributions of $p$ at two boundaries, the ScB problem is then defined as the minimization of the Kullback-Leibler (KL) divergence:
\begin{equation}
	\label{eqn:14}
	\setlength{\abovedisplayskip}{3pt}
	\setlength{\belowdisplayskip}{4pt}
	\min_{p\in\mathcal{P}_{[0,T]}}D_{\mathrm{KL}}(p\parallel p_{\mathrm{ref}}),\quad s.t. \  p_0=p_{\mathbf{X}},\ p_T=p_{\mathbf{Y}}, 
	\vspace{-3pt}
\end{equation}
where $\mathcal{P}_{[0, T]}$ denotes the space of all path measures over $[0, T]$, and $p_{\mathrm{ref}}$ is the reference path measure. For simplicity, $T$ is set to 1. When $p_{\mathrm{ref}}$ is defined by the same forward SDE as in Eq.~{\ref{eqn:11}}, the ScB problem is equivalent to a pair of forward-backward SDEs~{\cite{wang2021deep}}:
\begin{equation}
	\label{eqn:15}
	\setlength{\abovedisplayskip}{3pt}
	\setlength{\belowdisplayskip}{4pt}
	\mathrm{d}\boldsymbol{x}_t=[\boldsymbol{f}(\boldsymbol{x}_t,t)+g^2(t)\nabla\log\Psi_t(\boldsymbol{x}_t)]\mathrm{d}t+g(t)\mathrm{d}\boldsymbol{w}_t,\quad\boldsymbol{x}_0\sim p_\mathbf{X},
	\vspace{-3pt}
\end{equation}
\begin{equation}
	\label{eqn:16}
	\setlength{\abovedisplayskip}{3pt}
	\setlength{\belowdisplayskip}{4pt}
	\mathrm{d}\boldsymbol{x}_t=[\boldsymbol{f}(\boldsymbol{x}_t,t)-g^2(t)\nabla\log\widehat{\Psi}_t(\boldsymbol{x}_t)]\mathrm{d}t+g(t)\mathrm{d}\bar{\boldsymbol{w}}_t,\quad\boldsymbol{x}_T\sim p_{\mathbf{Y}},
	\vspace{-3pt}
\end{equation}
where $\boldsymbol{f}$ and $g$ are the same as that in the reference SDE. $\nabla\log\Psi_t(\boldsymbol{x}_t)$ and $\nabla\log\widehat{\Psi}_t(\boldsymbol{x}_t)$ are additional non-linear drift terms, described by the following coupled partial differential equations (PDEs):
\begin{equation}
	\label{eqn:17}
	\setlength{\abovedisplayskip}{3pt}
	\setlength{\belowdisplayskip}{4pt}
	\frac{\partial \Psi}{\partial t} = -\nabla_{\boldsymbol{x}}\Psi^{\mathsf{T}}\boldsymbol{f} - \frac{1}{2}\text{Tr}\left(g^{2}\nabla^{2}_{\boldsymbol{x}}\Psi\right),
	\vspace{-3pt}
\end{equation}
\begin{equation}
	\label{eqn:18}
	\setlength{\abovedisplayskip}{3pt}
	\setlength{\belowdisplayskip}{4pt}
	\frac{\partial \widehat{\Psi}}{\partial t} = -\nabla_{\boldsymbol{x}}\cdot\left(\widehat{\Psi}\boldsymbol{f}\right) + \frac{1}{2}\text{Tr}\left(g^{2}\nabla^{2}_{\boldsymbol{x}}\widehat{\Psi}\right).
	\vspace{-3pt}
\end{equation}

For any time $t\in[0, T]$, the marginal distribution $p_{t}$ should satisfy $p_{t} = \Psi_{t}\widehat{\Psi}_{t}$. In general, the ScB problem is not fully tractable, and closed-form solutions exist only for strict constrained families of $p_{\text{ref}}$~{\cite{bunne2023schrodinger}}.  Following~{\cite{chen2023schrodinger}}, we consider a simplified scenario where both boundaries are corrupted by small Gaussian noise, and the reference drift is linear, \emph{i.e.}, $\boldsymbol{f}\left(\boldsymbol{x}_{t}, t\right) = f\left(t\right)\boldsymbol{x}_{t}$. Two boundaries can be parameterized as $p_{\mathbf{X}}\sim\mathcal{N}_{\mathbb{C}}\left(\mathbf{X}, \epsilon_{0}^{2}\boldsymbol{I}\right)$ and $p_{\mathbf{Y}}\sim\mathcal{N}_{\mathbb{C}}\left(\mathbf{Y}, \epsilon_{1}^{2}\boldsymbol{I}\right)$. And we assume $\epsilon_{1} = e^{\int_{0}^{1}f\left(\tau\right)d\tau}\epsilon_{0}$. As $\epsilon_{0}\to 0$, the ScB problem becomes fully tractable, and the solutions of $\Psi_{t}$ and $\widehat{\Psi}_{t}$ are given by:
\begin{equation}
	\label{eqn:19}
	\setlength{\abovedisplayskip}{3pt}
	\setlength{\belowdisplayskip}{4pt}
	\widehat{\Psi}_{t} = \mathcal{N}_{\mathbb{C}}\left(\alpha_{t}\mathbf{X}, \alpha_{t}^{2}\sigma_{t}^{2}\boldsymbol{I}\right),\Psi_{t}=\mathcal{N}_{\mathbb{C}}\left(\bar{\alpha}_{t}\mathbf{Y}, \alpha_{t}^{2}\bar{\sigma}_{t}^{2}\boldsymbol{I}\right),
	\vspace{-3pt}
\end{equation}
where $\alpha_{t}=e^{\int_{0}^{t}f\left(\tau\right)d\tau}$, $\bar{\alpha}_{t}=e^{-\int_{t}^{1}f\left(\tau\right)d\tau}$, $\sigma_{t}^{2}=\int_{0}^{t}\frac{g^{2}\left(\tau\right)}{\alpha_{\tau}^{2}}d\tau$, $\bar{\sigma}_{t}^{2}=\int_{t}^{1}\frac{g^{2}\left(\tau\right)}{\alpha_{\tau}^{2}}d\tau$. As a result, the marginal distribution of the ScB is given by:
\begin{equation}
	\label{eqn:20}
	\setlength{\abovedisplayskip}{3pt}
	\setlength{\belowdisplayskip}{4pt}
	p_{t} = \Psi_{t}\widehat{\Psi}_{t} = \mathcal{N}\left(\frac{\alpha_{t}\bar{\sigma}_{t}^{2}\mathbf{X} + \bar{\alpha}_{t}\sigma_{t}^{2}\mathbf{Y}}{\sigma_{1}^{2}}, \frac{\alpha_{t}^{2}\bar{\sigma}_{t}^{2}\sigma_{t}^{2}}{\sigma_{1}^{2}}\boldsymbol{I}\right), 
\end{equation}
where $\sigma_{1}^{2} = \sigma_{t}^{2} + \bar{\sigma}_{t}^{2}$. As noted in~{\cite{chen2023schrodinger}}, the choice of noise schedules significantly impacts performance, and here we consider three types of schedules: variance-preserving (VP), variance-exploding (VE), and gmax, which are detailed in Table~{\ref{tbl:noiseschedules}}.
\begin{algorithm}[t]
	\caption{BridgeVoC Training Procedure}
	\label{alg:bridgevoc_training}
	\begin{algorithmic}[1]
		\Require{Mel-spectrum $\mathbf{X}^{\text{mel}}$, target spectrum $\mathbf{X}$}
		\State Initialize Range-space Surrogation:
		\State \Indent $\mathbf{Y} \gets \mathcal{A}^{\dagger}\exp\left(\mathbf{X}^{\text{mel}}\right)\exp\left(\mathbf{\Phi}_{\text{init}}\right)$ 
		\While{Training not converged}
		\State Sample Timestep and Noise:
		\State \Indent $t\sim\mathcal{U}(0, 1)$, $\boldsymbol{\epsilon} \sim \mathcal{N}(0, I)$
		\State Parameterize Input:
		\State \Indent $\boldsymbol{x}_t\gets\text{Parametrize via Eq.~{\ref{eqn:20}}}$ \Comment{Sample $\boldsymbol{x}_t$}
		\State Estimate:
		\State \Indent $\tilde{\mathbf{X}} \gets \mathcal{B}_{\theta}\left(\boldsymbol{x}_t, \mathbf{Y}, t\right)$, $\boldsymbol{\tilde{s}} \gets \text{iSTFT}(\tilde{\mathbf{X}})$
		\State Calculate data prediction loss $\mathcal{L}_{data}$ following Eq.~{\ref{eqn:29}}.
		\State Calculate Mel loss $\mathcal{L}_{mel}$ following Eq.~{\ref{eqn:30}}.
		\State Calculate adversarial loss $\mathcal{L}_{g}$ following Eq.~{\ref{eqn:31}}.
		\State Calculate feature matching loss $\mathcal{L}_{fm}$ following Eq.~{\ref{eqn:32}}.
		\State Update Diffusion Model:
		\State \Indent $\theta \gets \theta - \eta \cdot \nabla_{\theta} \mathcal{L}_{G}$ \Comment{$\mathcal{L}_{G}$ is defined in Eq.~{\ref{eqn:33}}}
		\EndWhile
	\end{algorithmic}
\end{algorithm}

\begin{table*}[t]
	\centering
	\caption{Reverse SDE and ODE samplers for the proposed BridgeVoC. $t\in\left[0, \tau\right)$.}
	\label{tbl:sb_samplers}
	\resizebox{0.98\textwidth}{!}{
		\begin{tabular}{cc}
			\hline\hline
			\multicolumn{1}{c}{SDE sampler} & \multicolumn{1}{c}{ODE sampler} \\
			\hline 
			$\displaystyle \boldsymbol{x}_t = \frac{\alpha_t \sigma_t^2}{\alpha_\tau \sigma_\tau^2} \boldsymbol{x}_\tau + \alpha_t \left( 1 - \frac{\sigma_t^2}{\sigma_\tau^2} \right) 
			\mathcal{B}_{\theta}\left(\boldsymbol{x}_{\tau}, \mathbf{Y}, t\right) + \alpha_t \sigma_t \sqrt{1 - \frac{\sigma_t^2}{\sigma_\tau^2}} \boldsymbol{\epsilon}$ 
			& 
			$\displaystyle \boldsymbol{x}_t = \frac{\alpha_t \sigma_t \bar{\sigma}_t}{\alpha_\tau \sigma_\tau \bar{\sigma}_\tau} \boldsymbol{x}_\tau + \frac{\alpha_t}{\bar{\sigma}_T^2} \left( \bar{\sigma}_t^2 - \frac{\bar{\sigma}_\tau \sigma_t \bar{\sigma}_t}{\sigma_\tau} \right) \mathcal{B}_{\theta}\left(\boldsymbol{x}_{\tau}, \mathbf{Y}, t\right) + \frac{\alpha_t}{\alpha_T \sigma_T^2} \left( \sigma_t^2 - \frac{\sigma_\tau \sigma_t \bar{\sigma}_t}{\sigma_\tau} \right) \mathbf{Y}$ \\
			\hline\hline
	\end{tabular}}
	\vspace{-6pt}
\end{table*}
\setcounter{figure}{4}
\begin{figure*}
	\centering
	\vspace{0pt}
	\includegraphics[width=0.85\textwidth]{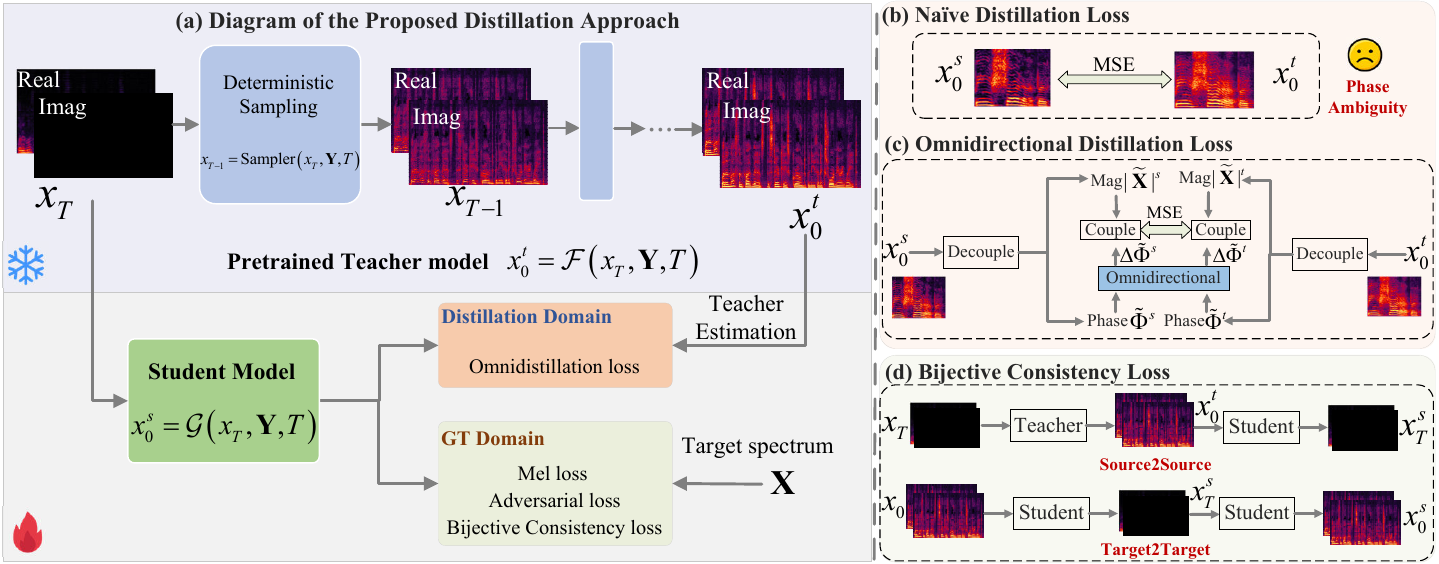}
	\vspace{-2pt}
	\caption{Diagram of the proposed single-step distillation scheme. (a) The single-step student model learns the deterministic mapping function from the teacher diffusion model, which involves two types of losses: distillation-related and ground-truth (GT)-related. (b) Detail of the na\"ive distillation loss. (c) Detail of the proposed omnidirectional distillation loss. (d) Detail of the bijective consistency loss, including two bijective mappings, namely source-to-source (S2S) and target-to-target (T2T).}
	\label{fig:distill}
	\vspace{-10pt}
\end{figure*}
\vspace{-8pt}
\subsection{Network Structure}\label{sec:network-structure}
\vspace{-2pt}
With the tractable solutions for $\Psi_{t}$ and $\widehat{\Psi}_{t}$ in Eq.~{\ref{eqn:19}}, we can iteratively restore the target spectrum by following the reverse process in Eqs.~{\ref{eqn:15}}-{\ref{eqn:16}}. Since $\mathbf{X}$ is involved in Eq.~{\ref{eqn:19}}, a neural network can be directly employed to estimate the target spectrum $\mathbf{X}$ given the intermediate state $\boldsymbol{x}_{t}$. In our conference paper~{\cite{tong2025bridge}}, we instantiate it with NCSN++~{\cite{song2020score}}, a classical U-Net architecture, whose structure is shown in Fig.~{\ref{fig:ncsn_system}}. NCSN++ includes a main feature extraction path of the multi-resolution U-Net structure (see bottom) and a progressive growing of the input (see top). The former is to gradually compress and recover the feature map, and the latter aims to project the feature space into the target space. Residual blocks are adopted between consecutive up/down layers. The attention block is inserted in the bottleneck. The input is a concatenation of the noisy state $\boldsymbol{x}_{t}$ and condition $\mathbf{Y}$, as well as the timestep $t$. For additional details, readers are referred to~{\cite{song2020score}}.

Despite its widespread use in generative tasks, NCSN++ has intrinsic limitations for audio generation. First, NCSN++ was originally designed for image tasks, where downsampling and upsamplings operations operate on fixed image width/height dimensions. Audio signals, by contrast, have variable lengths, leading to dynamic framing property. This mismatch makes NCSN++ cumbersome to train and prone to information loss from excessive downsampling. Second, the T-F spectrum exhibits distinct patterns across frequency regions, but NCSN++ ignores this hierarchical prior characteristic, often resulting in acoustic distortions.

To address these limitations, we propose a convolution-based subband-aware diffusion network dubbed BCD for vocoders, which explicitly incorporates subband modeling. As shown in Fig.~{\ref{fig:framework}}(a), BCD consists of three parts: Convolutional-Style Subband-Division (CSBD), Large-Kernel Convolutional Attention Module (LKCAM), and Convolutional-Style Subband-Merge (CSBM). Detailed illustrations are provided below.
\subsubsection{CSBD}\label{sec:cbdm}
The CSBD structure is detailed in Fig.~{\ref{fig:framework}}(c). Similar to NCSN++, we first concatenate $\boldsymbol{x}_{t}$ and $\mathbf{Y}$ along the channel axis to form the input:
\begin{equation}
	\label{eqn:21}
	\setlength{\abovedisplayskip}{3pt}
	\setlength{\belowdisplayskip}{4pt}
	\mathbf{I}_{t} = \text{Concat}\left(\boldsymbol{x}_{t}, \mathbf{Y}\right)\in\mathbb{R}^{4\times F\times L}.
	\vspace{-3pt}
\end{equation}

Note that the complex-valued tensor is converted into the real-valued format by concatenating the real and imaginary (RI) components along the channel axis. Considering that harmonic components mainly lie in the low- and mid-frequency regions, we adopt an uneven subband division strategy. Specifically, the input spectrum is split into $R$ regions. For the $r$-th region, we apply a separate Conv2d, followed by layer-normalization (LN), with kernel size and stride being $\left\{\left(k_{f}^{e}, k_{l}^{e}\right), \left(s_{f}^{e}, s_{l}^{e}\right)\right\}$, respectively:
\begin{equation}
	\label{eqn:22}
	\setlength{\abovedisplayskip}{3pt}
	\setlength{\belowdisplayskip}{4pt}
	\mathbf{F}_{in,r} = \text{LN}\left(\text{Conv2d}\left(\mathbf{I}_{t,r}\right)\right)\in\mathbb{R}^{C\times N_{r}\times L},
	\vspace{-3pt}
\end{equation}
where the subscript $\left(\cdot\right)_{r}$ denotes the region index, $C$ is the number of output channels, and $N_{r}$ is the number of compressed subbands for the $r$-th region. We  then concatenate the encoded representations of all regions to form the unified subband feature tensor as:
\begin{equation}
	\label{eqn:23}
	\setlength{\abovedisplayskip}{3pt}
	\setlength{\belowdisplayskip}{4pt}
	\mathbf{F}_{in} = \text{Concat}\left(\mathbf{F}_{in,1},\cdots,\mathbf{F}_{in,R}\right)\in\mathbb{R}^{C\times N\times L},
	\vspace{-3pt}
\end{equation}
where $N=\sum_{r=1}^{R}N_{r}$ is the total number of compressed subbands. To control the encoded feature stream across different diffusion timesteps, we modulate $\mathbf{F}_{in}$ as:
\begin{equation}
	\label{eqn:24}
	\setlength{\abovedisplayskip}{3pt}
	\setlength{\belowdisplayskip}{4pt}
	\mathbf{P} = \left(1 + \gamma_{in,t}\right)\mathbf{F}_{in} + \beta_{in,t},
	\vspace{-3pt}
\end{equation}   
where $\left\{\gamma_{in, t}, \beta_{in, t}\right\}$ are scale and shift parameters, generated from the timestep embedding $\mathbf{E}_{t}$.
\begin{figure}[t]
	\centering
	\vspace{0pt}
	\includegraphics[width=0.37\textwidth]{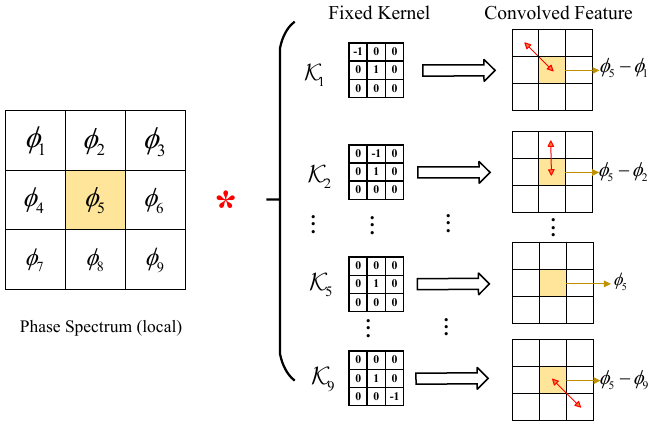}
	\vspace{-2pt}
	\caption{Diagram of the omnidirectional phase operation.}
	\label{fig:omni}
	\vspace{-0.5cm}
\end{figure}
\begin{figure}[t]
	\centering
	\vspace{0pt}
	\includegraphics[width=0.94\columnwidth]{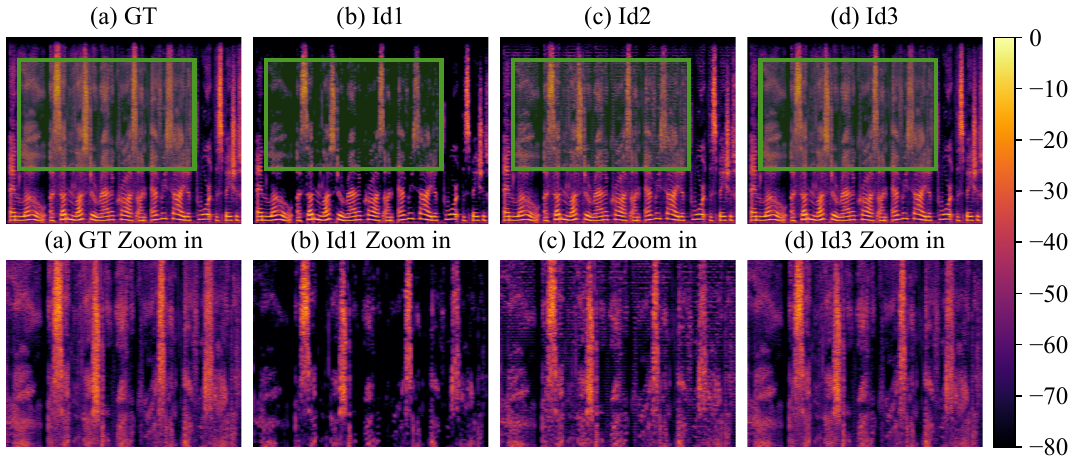}
	\vspace{-6pt}
	\caption{Spectral visualizations with different loss configurations. The audio file is taken from the LibriTTS \textit{dev-clean} set.}
	\label{fig:ablation_loss_visualization}
	\vspace{-0.35cm}
\end{figure}
\subsubsection{CSBM}\label{sec:cbmm}
In contrast to CSBD, CSBM decodes the latent subband features back to the full-frequency target spectrum. Given the latent feature tensor $\mathbf{O}\in\mathbb{R}^{C\times N\times T}$ (output of LKCAM), it is first modulated by the timestep-aware embedding $\mathbf{E}_{t}$, and split into $R$ regions. For each region, it passes a point-wise Conv2d (PConv2d), LN, and Gaussian error linear unit (GELU). After that, a Transposed Conv2d is given to recover the original frequency resolution. The process can be given by:
\begin{equation}
	\label{eqn:25}
	\setlength{\abovedisplayskip}{3pt}
	\setlength{\belowdisplayskip}{4pt}
	\mathbf{F}_{out,r} = \left(1 + \gamma_{out,t}\right)\mathbf{O} + \beta_{out, t}, 
	\vspace{-3pt}
\end{equation}
\begin{equation}
	\label{eqn:26}
	\setlength{\abovedisplayskip}{3pt}
	\setlength{\belowdisplayskip}{4pt}
	\mathbf{K}_{r} = \text{TrConv2d}\left(\text{GELU}\left(\text{LN}\left(\text{PConv2d}\left(\mathbf{F}_{out,r}\right)\right)\right)\right),
	\vspace{-3pt}
\end{equation}
where $\left\{\gamma_{out, t}, \beta_{out, t}\right\}$ are the corresponding time-adaptive modulation parameters. Note that, $\mathbf{K}_{r}$ is finally split into real and imaginary parts, and we can obtain the RI target estimation by concatenating $\left\{\mathbf{K}_{1},\cdots,\mathbf{K}_{R}\right\}$ along the frequency axis.
\subsubsection{LKCAM}\label{sec:lkcam}
In the T-F domain, both inter-frame and inter-band relations are significant for spectrum estimation. On one hand, audio signals can exhibit strong sequential characteristics, and neighboring frames can be helpful for current frame estimation. On the other hand, recent success on bandwidth extension (BWE) has revealed the implicit dependencies across frequency regions~{\cite{liu2024audiosr}}. To model these planar relationships efficiently, we propose LKCAM, an efficient convolutional-style attention module with large kernels, as shown in Fig.~{\ref{fig:framework}}(d).  

Specifically, it consists of $P$ stacked LKCABs, each of which is detailed in Fig.~{\ref{fig:framework}}(f), comprising a Convolutional Attention Block (CAB) and a Convolutional Feedforward Network (ConvFFN). For the first part, inspired by~{\cite{hou2024conv2former}}, we replace traditional self-attention (SA) with a convolutional modulation layer. Concretely, for $p$-th LKCAB, given the input feature $\mathbf{H}^{\left(p-1\right)}\in\mathbb{R}^{C\times N\times L}$, instead of calculating the attention matrix $\mathbf{A}^{\left(p\right)}$ via self-similarity, here the attention branch works by adopting the large-kernel convolution operation:
\begin{equation}
	\label{eqn:27}
	\setlength{\abovedisplayskip}{3pt}
	\setlength{\belowdisplayskip}{4pt}
	\mathbf{A}^{\left(p\right)} = \text{LKDWConv2d}\left(\text{GELU}\left(\text{PConv2d}\left(\text{LN}\left(\mathbf{H}^{\left(p-1\right)}\right)\right)\right)\right),
	\vspace{-3pt}
\end{equation}
where $\text{LKDWConv2d}\left(\cdot\right)$ denotes a large-kernel depthwise Conv2d with kernel size being $\left(k_{f}, k_{l}\right)$ that captures both inter-frame and inter-band relationships. We compute the value tensor $\mathbf{V}^{\left(p\right)}$ via point-wise convolution (PConv2d). The attended feature can be calculated as:
\begin{equation}
	\label{eqn:28}
	\setlength{\abovedisplayskip}{3pt}
	\setlength{\belowdisplayskip}{4pt}
	\mathbf{Z}^{\left(p\right)} = \text{PConv2d}\left(\mathbf{A}^{\left(p\right)}\odot\mathbf{V}^{\left(p\right)}\right),
	\vspace{-3pt}
\end{equation}
where ``$\odot$'' denotes element-wise multiplication. To manage the diffusion steps, the AdaLN layers are usually adopted. However, it can exhibit excessive parameters. To this end, we adopt AdaLN-SOLA~{\cite{hai2024ezaudio}}, a lightweight variant that uses block-specific low-rank matrices for modulation, and can preserve performance while reducing parameters.

For ConvFFN, following~{\cite{zhou2023srformer}}, we insert a DWConv2d with residual connections to enhance high-frequency detail encoding. Besides, the AdaLN-SOLA is adopted for timestep-aware feature modulation. 

\textbf{Remarks}: We would like to highlight the distinctions between our method and RFWave~{\cite{liu2024rfwave}}, a recently proposed subband-based diffusion vocoder. First, RFWave falls into the N2D paradigm, where flow-matching is employed to generate the target starting from pure Gaussian noise. Conversely, we adopt the D2D regime, restoring the target from a degraded range-space spectral representation. This notably shortens the generation trajectory and reduces modeling difficulty. Second, an even subband splitting strategy is adopted in RFWave while we employ an uneven scheme, where finer division is applied in the low- and mid-frequency regions. This can better preserve the acoustic features. Third, in contrast to simply reusing the network structure from Vocos~{\cite{siuzdakvocos}} for diffusion, we consider the relationships in both inter-frame and inter-band, and elaborately devise a large-kernel convolution module to support efficient T-F modeling. Experiments in Sec.~{\ref{sec:ablations}} demonstrate that these proposed strategies can substantially improve the performance of waveform reconstruction with fewer reverse steps, \emph{e.g.}, 4 steps, and no additional fast sampling strategies are required. 
\vspace{-0.45cm}
\subsection{Training and Inference Procedure}\label{sec:training-and-inference-procedure}
\vspace{-2pt}
Following~{\cite{jukic2024schr}}, we adopt a data prediction loss. Let $\tilde{\mathbf{X}} = \mathcal{B}_{\theta}\left(\boldsymbol{x}_{t}, \mathbf{Y}, t\right)$ denotes the estimated target spectrum, where $\mathcal{B}_{\theta}\left(\cdot\right)$ is a mapping function instantiated by the data prediction network. The data prediction loss can be defined as:
\begin{equation}
	\label{eqn:29}
	\setlength{\abovedisplayskip}{3pt}
	\setlength{\belowdisplayskip}{4pt}
	\mathcal{L}_{data} = \frac{1}{FL}\sum_{f,l}\left\|\tilde{\mathbf{X}}_{f, l} - \mathbf{X}_{f, l}\right\|_{2}^{2}.
	\vspace{-3pt}
\end{equation}

To further facilitate the restoration of acoustic features, we incorporate the multi-resolution Mel loss:
\begin{equation}
	\label{eqn:30}
	\setlength{\abovedisplayskip}{3pt}
	\setlength{\belowdisplayskip}{4pt}
	\mathcal{L}_{mel} = \frac{1}{F_{m}L}\sum_{i=1}^{I	}\sum_{f,l}\left\|\tilde{\mathbf{X}}^{mel,i}_{f, l} - \mathbf{X}^{mel,i}_{f, l}\right\|_{1},
	\vspace{-3pt}
\end{equation}
where superscript $\left(\cdot\right)^{i}$ denotes the Mel scale index. Furthermore, when Mel loss is adopted, although notable metric improvements are achieved, some point-like artifacts are observed in mid- and high-frequency regions (see Fig.~{\ref{fig:ablation_loss_visualization}}). To this end, we additionally incorporate the adversarial loss~{\cite{wangdiffusion}}. To be specific, the multi-periodic discriminator (MPD)~{\cite{kong2020hifi}} and multi-resolution STFT discriminator (MRD)~{\cite{defossez2023high}} are utilized. For generator, the adversarial loss can be defined as:
\begin{equation}
	\label{eqn:31}
	\setlength{\abovedisplayskip}{3pt}
	\setlength{\belowdisplayskip}{4pt}
	\mathcal{L}_{g} = \frac{1}{M}\sum_{m=1}^{M}\max\left(0, 1-D_{m}\left(\boldsymbol{\tilde{s}}\right)\right).
	\vspace{-3pt}
\end{equation}

Besides, we also include a feature matching loss, given by:
\begin{equation}
	\label{eqn:32}
	\setlength{\abovedisplayskip}{3pt}
	\setlength{\belowdisplayskip}{4pt}
	\mathcal{L}_{fm} = \frac{1}{UM}\sum_{u,m}\left\|\mathbf{f}_{u}^{m}\left(\boldsymbol{\tilde{s}}\right) - \mathbf{f}_{u}^{m}\left(\boldsymbol{s}\right)\right\|_{1},
	\vspace{-3pt}
\end{equation}
where $\mathbf{f}_{u}^{m}\left(\cdot\right)$ denotes the $u$-th intermediate feature map of the $m$-th discriminator. The total generator loss can be defined as:
\begin{equation}
	\label{eqn:33}
	\setlength{\abovedisplayskip}{3pt}
	\setlength{\belowdisplayskip}{4pt}
	\mathcal{L}_{G} = \lambda_{data}\mathcal{L}_{data} + \lambda_{mel}\mathcal{L}_{mel} + \lambda_{g}\mathcal{L}_{g} + \lambda_{fm}\mathcal{L}_{fm},
	\vspace{-3pt}
\end{equation} 
where $\left\{\lambda_{data}, \lambda_{mel}, \lambda_{g}, \lambda_{fm}\right\}$ denote the weighting parameters. Detailed training procedure is presented in Algorithm~{\ref{alg:bridgevoc_training}}. 

For inference, following~{\cite{chen2023schrodinger}}, both SDE and ordinary differential equation (ODE) formulations are investigated. Starting from the initial state $\boldsymbol{x}_{T} = \mathbf{Y}$, we iterate the reverse process using the update rules in Table~{\ref{tbl:sb_samplers}} until $t=0$. The final output $\boldsymbol{x}_{0} = \mathbf{\tilde{X}}$ can be converted to a time-domain waveform $\boldsymbol{\tilde{s}}$ via inverse STFT.
\vspace{-0.25cm}
\subsection{Distill Single-step Generation}\label{sec:distilled-single-step}
\vspace{-3pt}
Processing efficiency is critical for deploying neural vocoders in real-time scenarios. While diffusion vocoders achieve high quality, their multi-step inference hinders speed. Therefore, we propose a distilled single-step diffusion model based on BridgeVoC, denoted, \textbf{BridgeVoC$^{\star}$}, in which a lightweight student model learns a deterministic mapping from a pre-trained teacher diffusion model, as shown in Fig.~{\ref{fig:distill}}(a). Given a pre-trained teacher diffusion model, we gradually reconstruct the target spectrum via multiple reverse steps, and the ODE solver is applied to guarantee deterministic generation. The reverse process of the teacher model is given by:
\begin{equation}
	\label{eqn:34}
	\setlength{\abovedisplayskip}{3pt}
	\setlength{\belowdisplayskip}{4pt}
	\boldsymbol{x}_{0}^{t} = \mathcal{G}_{tea}\left(\boldsymbol{x}_{T}, \mathbf{Y}, T\right),
\end{equation}
where superscript $\left(\cdot\right)^{t}$ denotes the estimation by the teacher model, and $\mathcal{G}_{tea}\left(\cdot\right)$ denotes the deterministic reverse mapping function of the teacher model. The student model aims to learn a single-step mapping. Given the single-step output of the student model as $\boldsymbol{x}_{0}^{s}$, a na\"ive distillation loss is defined as:
\begin{equation}
	\label{eqn:35}
	\setlength{\abovedisplayskip}{3pt}
	\setlength{\belowdisplayskip}{4pt}
	\mathcal{L}_{naivedistill} = \frac{1}{FL}\sum_{f,l}\left\|\boldsymbol{x}_{0}^{t} - \boldsymbol{x}_{0}^{s}\right\|_{2}^{2}.
	\vspace{-6pt}
\end{equation}

However, we observe severe spectral distortion with this loss (see Fig.~{\ref{fig:single_step_visualization}(b)(h)), which is attributed to \textit{information ambiguity induced by phase}, as shown in Fig.~{\ref{fig:distill}}(b). Concretely, the RI components can be regarded as the cosine and sine modulated form of magnitude spectrum by phase. In the vocoder task, estimating the instantaneous phase remains challenging due to the absence of a reference phase and the inherent phase-wrapping effect{\footnote{One may notice that the validation curves of instantaneous phase only fluctuates slightly throughout the entire training process. The tensorboard address: https://huggingface.co/Bakerbunker/FreeV\_Model\_Logs/tensorboard.}}. Consequently, phase misalignment can lead to ambiguity in the RI components between the estimation of teacher and student models in Eq.~{\ref{eqn:35}}, thus hindering the effective knowledge transfer.

Recently, some phase optimization methods have been proposed by incorporating differential operations along the time and frequency axes~{\cite{ai2023apnet}}, \emph{i.e.}, group delay (GD) and instantaneous frequency (IF), which exhibit better spectral structure. To this end, we propose an omnidirectional distillation loss, where we generalize the differential operation into all neighboring T-F bins, as shown in Fig.~{\ref{fig:distill}}(c). To be specific, $\left\{\boldsymbol{x}_{0}^{s}, \boldsymbol{x}_{0}^{t}\right\}$ are first decoupled into magnitude and phase components. For phase, we employ an omnidirectional operation to capture differential relations with adjacent T-F bins, whose diagram is shown in Fig.~{\ref{fig:omni}}. We elaborately devise nine $3\times 3$ kernels with fixed parameters, and a simple convolution operation is applied to the phase. The convolved phase is coupled with the magnitude to calculate the distillation loss. The whole process can be formulated as:
\begin{equation}
	\label{eqn:36}
	\setlength{\abovedisplayskip}{3pt}
	\setlength{\belowdisplayskip}{4pt}
	\Delta{\tilde{\Phi}^{s}} = \text{Omni}\left(\tilde{\Phi}^{s}\right), \Delta{\tilde{\Phi}^{t}} = \text{Omni}\left(\tilde{\Phi}^{t}\right)
	\vspace{-3pt}
\end{equation}
\begin{equation}
	\label{eqn:37}
	\setlength{\abovedisplayskip}{3pt}\setlength{\belowdisplayskip}{4pt}\Delta{\mathbf{\tilde{X}}^{s}} = \left|\tilde{\mathbf{X}}^{s}\right|\Delta{\tilde{\Phi}^{s}}, \Delta{\mathbf{\tilde{X}}^{t}} = \left|\tilde{\mathbf{X}}^{t}\right|\Delta{\tilde{\Phi}^{t}},
	\vspace{-3pt}
	\end{equation}
\begin{equation}
	\label{eqn:38}
	\setlength{\abovedisplayskip}{1pt}
	\setlength{\belowdisplayskip}{2pt}
	\mathcal{L}_{omnidistill} = \frac{1}{FL}\sum_{f,l}\left\|\Delta{\mathbf{\tilde{X}}^{t}} - \Delta{\mathbf{\tilde{X}}^{s}}\right\|_{2}^{2},
	\vspace{-6pt}
\end{equation}
where $\text{Omni}\left(\cdot\right)$ denotes omnidirectional operation on phase, $\Delta$ denotes the convoluted component. The introduction of omnidirectional operation can effectively alleviate the optimization hurdled by phase and facilitate the information distillation from the teacher to student model, as visualized in Fig.~{\ref{fig:single_step_visualization}}.

A limitation of the distillation loss is that the GT audio is not utilized in the training, thus restricting the upper-bound performance of the student model. To this end, we also incorporate the Mel loss and adversarial loss, whose definitions have been given in Sec.~{\ref{sec:training-and-inference-procedure}}. Besides, inspired by~{\cite{wang2024sinsr}}, we adopt a consistency-preservation loss to enforce the student model to learn the bijective mapping of source and target data, as shown in Fig.~{\ref{fig:distill}}(d). To be specific, given the source $\boldsymbol{x}_{T}$, it is first processed by the teacher model, then the output passes the student model to learn the inverse mapping, given by:
\begin{equation}
	\label{eqn:39}
	\setlength{\abovedisplayskip}{3pt}
	\setlength{\belowdisplayskip}{4pt}
	\mathcal{L}_{inverse} = \frac{1}{FL}\sum_{f,l}\left\|\mathcal{G}_{stu}\left(\mathcal{G}_{tea}\left(\boldsymbol{x}_{T},\mathbf{Y}, T\right),\mathbf{Y}, 0\right) - \boldsymbol{x}_{T}\right\|_{2}^{2},
\end{equation} 
where $\mathcal{G}_{stu}\left(\cdot\right)$ denotes the mapping function by the student model. For the target $\boldsymbol{x}_{0}$, it is first processed by the student model to obtain the degradation estimation, and then passes the student model again for target reconstruction, given by:
\begin{small}
\begin{equation}
	\label{eqn:40}
	\setlength{\abovedisplayskip}{3pt}
	\setlength{\belowdisplayskip}{4pt}
	\mathcal{L}_{gt} = \frac{1}{FL}\sum_{f,l}\left\|\mathcal{G}_{stu}\left(\text{Detach}\left(\mathcal{G}_{stu}\left(\boldsymbol{x}_{0},\mathbf{Y}, 0\right)\right),\mathbf{Y}, T\right) - \boldsymbol{x}_{0}\right\|_{2}^{2},
\end{equation}
\end{small}
where $\text{Detach}\left(\cdot\right)$ denotes the gradient-stop operation. The overall loss for distillation can be defined as:
\begin{align}
	\label{eqn:41}
	\setlength{\abovedisplayskip}{3pt}
	\setlength{\belowdisplayskip}{4pt}
	\mathcal{L}_{singlestep} &= \lambda_{distill}\mathcal{L}_{omnidistill} + \lambda_{mel}\mathcal{L}_{mel} + \lambda_{g}\mathcal{L}_{g}\nonumber \\ & + \lambda_{fm}\mathcal{L}_{fm} + \lambda_{inverse}\mathcal{L}_{inverse} + \lambda_{gt}\mathcal{L}_{gt},
	\vspace{-4pt}
\end{align}
where $\left\{\lambda_{distill},\lambda_{mel},\lambda_{g},\lambda_{fm},\lambda_{inverse},\lambda_{gt}\right\}$ denotes the corresponding weighting coefficients. Note that we instantiate the parameters of the student model from the teacher one to speed up the training{\footnote{We would like to acknowledge that there are also other single-step distillation methods, \emph{e.g.}, consistency distillation~{\cite{song2023consistency}}. We are not intended to explore the optimality, which is out of scope of this paper.}}. 
\renewcommand\arraystretch{0.85}
\begin{table*}[t]
	\centering
	\caption{Descriptions of the baseline methods on the spatialized LibriSpeech dataset.}
	\resizebox{0.82\textwidth}{!}{
		\begin{tabular}{cccccc}
			\hline\hline
			Ids &Methods &Type &Publish &Domain &Code Link\\
			\hline
			1 &HiFiGAN~{\cite{kong2020hifi}}   &GAN  &NeurIPS 2020 &T &https://github.com/jik876/hifi-gan\\
			2&iSTFTNet~{\cite{kaneko2022istftnet}}  &GAN  &ICASSP 2022 &T &https://github.com/rishikksh20/iSTFTNet-pytorch\\
			3 &Avocodo~{\cite{bak2023avocodo}}  &GAN &AAAI 2023  &T &https://github.com/ncsoft/avocodo\\
			4 &BigVGAN-base~{\cite{leebigvgan}}  &GAN &ICLR 2023  &T &https://github.com/NVIDIA/BigVGAN\\
			5 &BigVGAN~{\cite{leebigvgan}}  &GAN &ICLR 2023  &T &https://github.com/NVIDIA/BigVGAN\\
			6 &APNet~{\cite{ai2023apnet}}  &GAN &TASLP 2023  &T-F &https://github.com/YangAi520/APNet\\
			7 &APNet2~{\cite{du2023apnet2}}  &GAN &NCMMSC 2023 &T-F &https://github.com/redmist328/APNet2\\
			8 &Vocos~{\cite{siuzdakvocos}}  &GAN  &ICLR 2024 &T-F &https://github.com/gemelo-ai/vocos\\
			9 &FreGrad~{\cite{nguyen2024fregrad}}  &DDPM &ICASSP 2024  &T &https://github.com/kaistmm/fregrad\\
			10 &PriorGrad~{\cite{leepriorgrad}} &DDPM &ICLR 2022 &T &https://github.com/microsoft/NeuralSpeech/tree/master/PriorGrad-vocoder\\
			11 &PeriodWave~{\cite{lee2024periodwave}} &FM &ICLR 2025 &T &https://github.com/sh-lee-prml/PeriodWave\\
			12 &RFWave~{\cite{liu2024rfwave}} &FM &ICLR 2025 &T-F &https://github.com/bfs18/rfwave\\
			13 &WaveFM~{\cite{luo2025wavefm}} &FM &NAACL 2025 &T &https://github.com/luotianze666/WaveFM\\
			\hline\hline
	\end{tabular}}
	\label{tbl:descriptions-baselines}
	\vspace{-6pt}
\end{table*}
\renewcommand\arraystretch{0.8}
\begin{table*}[t]
	\centering
	\caption{Ablation studies conducted on the LibriTTS benchmark in terms of noise scheduler, loss configurations, and inference samplers. ``w/i Pred.'' denotes whether the data prediction loss in Eq.~{\ref{eqn:29}} is adopted. ``w/i Mel'' and ``w/i GAN'' denote whether the Mel loss in Eq.~{\ref{eqn:30}} and adversarial loss in Eqs.~{\ref{eqn:31}}-{\ref{eqn:32}} are adopted, respectively. ``$\downarrow$'' denotes the lower the better and ``$\uparrow$'' denotes the opposite situation. The proposed BCD serves as the network, and the number of function evaluations $\text{NFE}=4$. The best and second-best performances are highlighted in $\textbf{bold}$ and \underline{underlined}, respectively.}
	\vspace{-6pt}
	\resizebox{0.85\textwidth}{!}{
		\label{tab:ablations}
		\begin{tabular}{ccccccccccccc}
			\toprule
			&\textbf{Noise} &\multicolumn{3}{c}{\textbf{Loss Setups}} &\multirow{2}*{\textbf{Sampler}} &\multirow{2}*{M-STFT$\downarrow$} &\multirow{2}*{PESQ$\uparrow$} &\multirow{2}*{MCD$\downarrow$}  &Periodicity$\downarrow$ &\multirow{2}*{V/UV F1$\uparrow$} &\multirow{2}*{UTMOS$\uparrow$} &\multirow{2}*{VISQOL$\uparrow$}\\
			\cmidrule{3-5}
			\multirow{-2}*{\textbf{Ids}}&\textbf{Scheduler} &\textbf{w/i Pred.} &\textbf{w/i Mel} &\textbf{w/i GAN} & & & & &RMSE & & &\\
			\midrule
			1 &\multirow{4}*{gmax} &\Checkmark &\XSolidBrush &\XSolidBrush &SDE &2.400 &3.583 &2.517 &0.206 &0.887 &3.457 &3.544\\
			2 & &\Checkmark &\Checkmark &\XSolidBrush &SDE &0.775 &\underline{4.386} &\underline{1.680} &0.0664 &0.973 &\underline{3.738} &4.933\\
			3 & &\Checkmark &\Checkmark &\Checkmark &SDE &\textbf{0.708} &\textbf{4.427} &\textbf{1.534} &\textbf{0.061} &\textbf{0.976} &\textbf{3.746} &4.943\\
			4 & &\Checkmark &\Checkmark &\Checkmark &ODE &0.851 &4.233 &1.730 &\underline{0.064} &0.974 &3.735 &4.845\\
			\midrule
			5 &\multirow{2}*{VE} &\Checkmark &\Checkmark &\Checkmark &SDE &\underline{0.722} &4.364 &1.760 &0.068 &\underline{0.974} &3.486 &\textbf{4.957}\\
			6 & &\Checkmark &\Checkmark &\Checkmark &ODE &0.735 &4.379 &1.730 &0.068 &0.973 &3.567 &\underline{4.945}\\
			\midrule
			7 &\multirow{2}*{VP} &\Checkmark &\Checkmark &\Checkmark &SDE &0.741 &4.361 &1.799 &0.070 &0.971 &3.653 &4.914\\
			8 & &\Checkmark &\Checkmark &\Checkmark &ODE &- &- &- &- &- &- &-\\
			\bottomrule
	\end{tabular}}
	\vspace{-8pt}
\end{table*}
\renewcommand\arraystretch{0.90}
\begin{table}[t]
	\centering
	\caption{Ablation studies on network structure. ``M'' denotes 1 million parameters. ``$a\%\downarrow$'' denotes the percentage reduction in computational complexity of BCD compared to the 64.8M version NCSN++ under the same NFE.}
	\vspace{-6pt}
	\resizebox{\columnwidth}{!}{
		\label{tbl:ablations-network}
		\begin{tabular}{ccccccc}
			\toprule
			\multirow{2}*{\textbf{Network}} &\multirow{2}*{\textbf{NFE}} &\textbf{\#Param} &\textbf{MACs} &\multirow{2}*{M-STFT$\downarrow$} &\multirow{2}*{PESQ$\uparrow$}  &\multirow{2}*{UTMOS$\uparrow$}\\
			& &\textbf{(M)} &\textbf{(Giga/5s)} & & &\\
			\midrule 
			\multirow{4}*{NCSN++} &4 &\multirow{2}*{16.8} &981 &1.016 &4.376 &3.604\\
			&8 & &1962 &1.019 &4.419 &3.628\\
			\cmidrule{2-7}
			&4 &\multirow{2}*{64.8} &3904 &1.007 &4.423 &3.705 \\
			&8 & &7808 &1.010 &4.447 &3.734 \\
			\midrule
			\multirow{2}*{BCD(Pro.)} &4 &\multirow{2}*{7.65} &171(95.62\%$\downarrow$) &0.708 &4.427 &3.746 \\
			&8 & &343(95.61\%$\downarrow$) &0.689 &4.436 &3.765 \\
			\bottomrule
	\end{tabular}}
	\vspace{-0.3cm}
\end{table}
\renewcommand\arraystretch{0.85}
\begin{table}[t]
	\centering
	\caption{Ablation studies on basic modeling blocks.}
	\vspace{-6pt}
	\resizebox{0.90\columnwidth}{!}{
		\label{tbl:ablations-basic-architecture}
		\begin{tabular}{cccccc}
			\toprule
			\textbf{Basic Type} &M-STFT$\downarrow$ &PESQ$\uparrow$ &MCD$\downarrow$ &UTMOS$\uparrow$ &VISQOL$\uparrow$\\
			\midrule 
			Set-A &0.771 &4.250 &2.010 &3.461 &4.874\\
			Set-B &0.786 &4.241 &2.079 &3.503 &4.884\\
			Ours &\textbf{0.708} &\textbf{4.427} &\textbf{1.534} &\textbf{3.746} &\textbf{4.963}\\
			\bottomrule
	\end{tabular}}
	\vspace{-0.3cm}
\end{table}
\renewcommand\arraystretch{0.80}
\begin{table*}[t]
	\centering
	\caption{Ablation studies on the number of subbands. Four steps are utilized for inference.}
	\LARGE
	\resizebox{0.95\textwidth}{!}{
		\begin{tabular}{cccccccccccccc}
			\toprule
			\textbf{Subband} &\multirow{2}*{\textbf{w/i Even}} &\textbf{\#Pram.} &\multicolumn{3}{c}{\textbf{\#MACs (Giga/5s)}} &\multirow{2}*{M-STFT$\downarrow$} &\multirow{2}*{PESQ$\uparrow$} &\multirow{2}*{MCD$\downarrow$} &V/UV$\uparrow$ &Periodicity$\downarrow$  &\multirow{2}*{UTMOS $\uparrow$} &\multirow{2}*{VISQOL$\uparrow$}  &\multirow{2}*{WER$\downarrow$}\\
			\cmidrule{4-6}
			\textbf{Number} $\textbf{N}$ & &\textbf{(M)}  &\textbf{CSBD} &\textbf{LKCAM} &\textbf{CSBM} & & & &F1 &RMSE  & & & \\
			\midrule
			6 &\XSolidBrush &8.63 &2.95 &32.44 &111.16 &0.850 &4.191 &2.286 &0.961 &0.091 &3.579 &4.819 &6.84\\
			12 &\XSolidBrush &7.98  &2.96 &64.88 &60.04 &0.774 &4.309 &1.900 &0.961 &0.088 &3.666 &4.889 &6.61\\
			24 &\XSolidBrush &7.65 &2.96 &129.76 &38.96 &\underline{0.708} &\underline{4.427} &\underline{1.534} &\underline{0.976} &\underline{0.061} &\underline{3.746} &\textbf{4.963} &6.64\\
			24 &\Checkmark &7.58 &2.91 &129.76 &32.24 &0.786 &4.320 &1.850 &0.973 &0.069 &3.683 &4.937 &\underline{6.56}\\
			48 &\XSolidBrush &7.48 &2.97 &259.48 &37.32 &\textbf{0.689} &\textbf{4.465} &\textbf{1.410} &\textbf{0.980} &\textbf{0.054} &\textbf{3.751} &\underline{4.958} &\textbf{6.26}\\
			\bottomrule
	\end{tabular}}
	\label{tbl:objective-different-subband}
	\vspace{-6pt}
\end{table*}
\begin{figure*}[t]
	\centering
	\vspace{0pt}
	\includegraphics[width=0.78\textwidth]{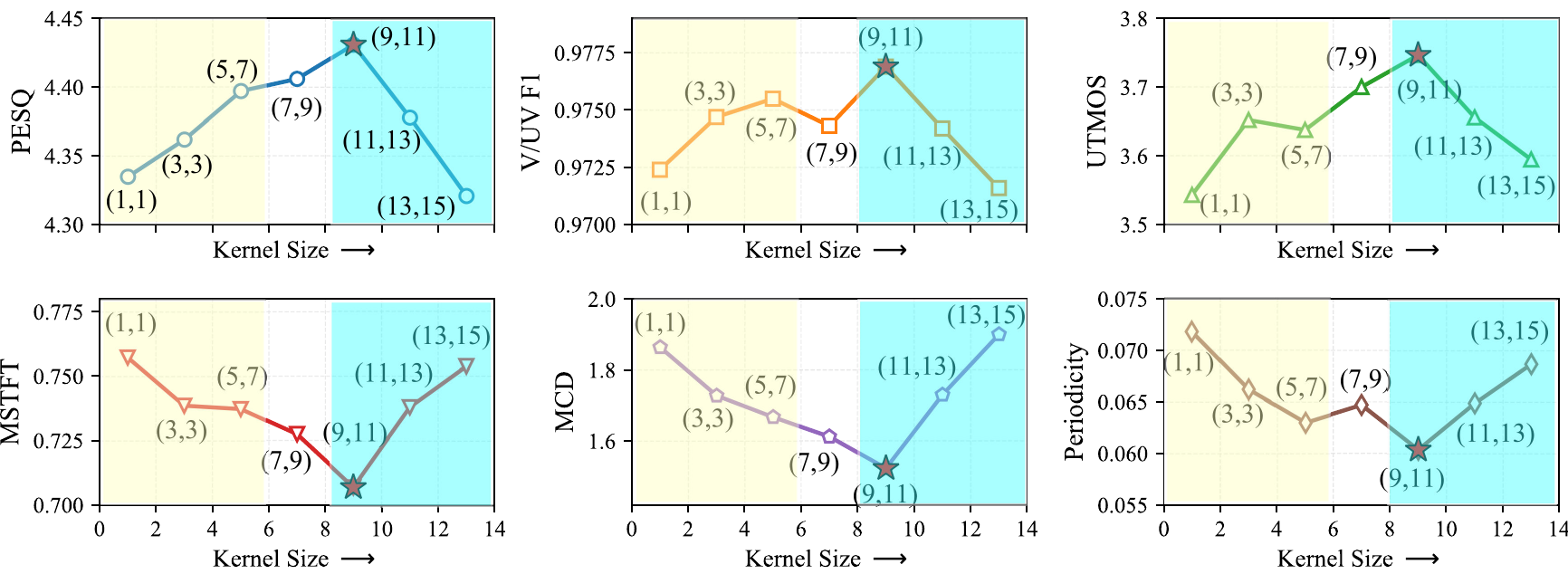}
	\vspace{-2pt}
	\caption{Metric scores under different $\left\{k_{f}, k_{t}\right\}$ settings within the LKCAB. For the first row, higher score indicates better performance. For the second row, lower score refers to better performance.}
	\label{fig:ablation_kernel_size}
	\vspace{-0.4cm}
\end{figure*}
\vspace{-6pt}
\section{Experimental Setups}\label{sec:experimental-setups}
\vspace{-2pt}
\subsection{Datasets}\label{sec:datasets}\vspace{-2pt}
In this paper, two benchmarks are adopted for model training: LJSpeech~{\cite{ljspeech17}} and LibriTTS~{\cite{zen19_interspeech}}. The LJSpeech dataset consists of 13,100 clean speech clips from a single female speaker, sampled at 22.05 kHz. Following the data split in the open-sourced VITS repository{\footnote{https://github.com/jaywalnut310/vits/tree/main/filelists}}, we use  $\left\{12500, 100, 500\right\}$ clips for training, validation and testings, respectively. The LibriTTS dataset features diverse acoustic environment and speaker characteristics, rendering it more challenging for waveform reconstruction. The original sampling rate is 24 kHz. Following~{\cite{leebigvgan}}, we use all the clips from $\left\{\textit{train-clean-100}, \textit{train-clean-360}, \textit{train-other-500}\right\}$ for training, and $\textit{val-full}$ for validation. 208 clips are selected from $\left\{\textit{dev-clean}, \textit{dev-other}\right\}$ as the test set, which is same to that of the open-sourced repository of BigVGAN{\footnote{https://github.com/NVIDIA/BigVGAN/tree/main/filelists}}.

To access generalization, we use four out-of-distribution (OOD) datasets: EARS~{\cite{richter2024ears}}, an expressive anechoic English dataset with diverse emotional styles; AISHELL3~{\cite{shi2020aishell}}, a Mandarin TTS dataset. We randomly select 200 clips from each of the two datasets for metric evaluations. We also incorporate MUSDB18~{\cite{rafii2017musdb18}} and FSD50K~{\cite{fonseca2021fsd50k}} for subjective evaluations and result visualizations, which are common music and sound effect datasets, respectively.
\vspace{-12pt}
\subsection{Miscellaneous Configurations}\label{sec:configurations}
\subsubsection{Mel Configurations}\label{sec:mel-configuration}
For the LJSpeech benchmark, the target sampling rate is set to 22.05 kHz, with 80 Mel-spectrum bins ($F_{m}=80$) and an upper-bound frequency ($f_{max}$) of 8 kHz. For the LibriTTS benchmark, the target sampling rate is 24 kHz, with $F_{m}$ being 100 and $f_{max}$ set to 12 kHz. For both datasets, the window size is 1024, with 25\% shift between adjacent frames, and 1024-point FFT is adopted, resulting in 513-dimension representation along the frequency axis. Following~{\cite{peer2023diffphase}}, the signal normalization is applied, and we choose $\left\{0.5, 0.33\right\}$ for spectrum exponential compression and gain scaling, respectively.
\begin{figure*}[t]
	\centering
	\vspace{0pt}
	\includegraphics[width=0.92\textwidth]{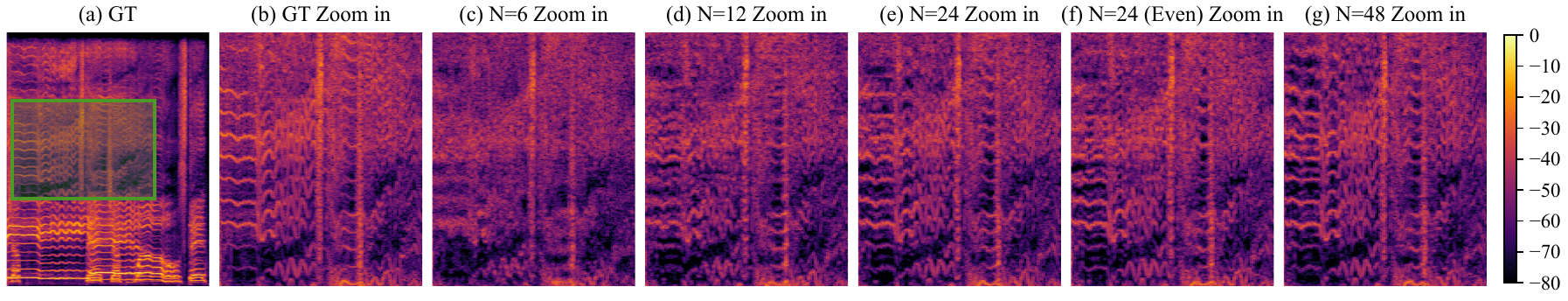}
	\vspace{-7pt}
	\caption{Spectral visualizations \emph{w.r.t.} different number of subbands. The audio file is a vocal clip from the MUSDB18 dataset.}
	\label{fig:ablation_nbands_visualization}
	\vspace{-0.35cm}
\end{figure*}
\subsubsection{Network Configurations}\label{sec:network-configurations}
For the proposed BCD, the CSBD divides the frequency axis into 3 regions ($R = 3$), with kernel size of $\left\{\left(12, 3\right), \left(24, 3\right), \left(44, 3\right)\right\}$ along the frequency and time axes, and the corresponding stride sizes of $\left\{\left(12, 1\right), \left(24, 1\right), \left(44, 1\right)\right\}$, yielding $N= 24$ total subbands. The CSBM uses symmetric kernel and stride settings to reverse the subband decoding and recover the full-frequency target spectrum. For the LKCAM, the encoded feature channel size $C$ is fixed at 256 by default, with $P=8$ LKCABs stacked to model 
long-range spectro-temporal dependencies. As a result, the number of trainable parameters for BCD is 7.97 M. For comparison, for NCSN++, two variants are considered, where the number of feature channels begins at 96 and 128, leading to 16.18 M and 64.54 M parameters, respectively.
\subsubsection{Training Configurations}\label{sec:training-configurations}
For standard diffusion training of BridgeVoC, a batch size of 8 and frame size of 128 are used, with an initial learning rate of 3e-4. The AdamW optimizer is employed with $\left\{\beta_{1}, \beta_{2}\right\}$ being set to $\left\{0.8, 0.99\right\}$. The generator and discriminators are updated for 1 M steps, with loss weighting hyper-parameters in Eq.~{\ref{eqn:33}}, $\left\{\lambda_{data}, \lambda_{mel}, \lambda_{g}, \lambda_{fm}\right\}$ are set to $\left\{1.0, 0.1, 20.0, 20.0\right\}$, respectively, to balance reconstruction and perceptual quality. For single-step model distillation, 16 deterministic reverse steps are empirically adopted for the teacher model, and the training setup for the student model is similar but with fewer updates (10 k steps) and the learning rate is initialized at 8e-5. The weighting hyper-parameters in Eq.~{\ref{eqn:41}} $\left\{\lambda_{distill},\lambda_{mel},\lambda_{g},\lambda_{fm},\lambda_{inverse},\lambda_{gt}\right\}$ are set to $\left\{1.0, 0.1, 20.0, 20.0, 1.0, 1.0\right\}$, respectively. The model is trained based on the Pytorch 2.3.0 framework.
\subsubsection{Baseline Models}\label{sec:baseline-models}
13 advanced vocoder baselines are included for comprehensive comparison, covering 8 GAN-based models and 5 diffusion-based models. More detailed information is provided in Table~{\ref{tbl:descriptions-baselines}}.
\subsubsection{Metrics}\label{sec:metrics}
Both objective and subjective metrics are used to evaluate performance. For the former, eight metrics are employed: (1) Multi-resolution STFT (M-STFT)~{\cite{yamamoto2020parallel}}, which quantifies spectral distance across multiple T-F resolutions. (2) Wide-band version of perceptual evaluation of speech quality (PESQ)~{\cite{rec2005p}}, which measures the objective speech quality. (3) Mel-cepstral distortion (MCD)~{\cite{kubichek1993mel}}, which computes the Mel difference via dynamic time wrapping (DTW). (4) Periodicity RMSE and V/UV F1 score~{\cite{morrisonchunked}}, which are viewed as major artifacts for non-autoregressive neural vocoders. (5) UTMOS~{\cite{saeki2022utmos}}, a non-intrusive automatic MOS predictor that evaluates quality without requiring a reference signal. (6) VISQOL~{\cite{hines2015visqol}}, an intrusive metric that scores spectro-temporal similarity between generated and reference signals. (7) Word error rate (WER) and Character error rate (CER), computed using OpenAI's Whisper pre-trained model. The former is based on Base and the latter is based on Large{\footnote{https://github.com/openai/whisper}}.

For subjective evaluations, MOS testing and AB testing are conducted. Twenty-four participants majoring audio/speech engineering from China are recruited, with each receiving 200 RMB upon completing the listening test. In MOS testing, participants rate each audio clip on a $[1, 5]$ scale (0.5 intervals) based on overall similarity to the reference. In AB testing. they select the clip with better quality from a pair, or choose ``Equal'' if no preference exists. The order of clips is randomly shuffled for each participant, to mitigate rating bias.
\begin{figure}[t]
	\centering
	\vspace{0pt}
	\includegraphics[width=0.44\textwidth]{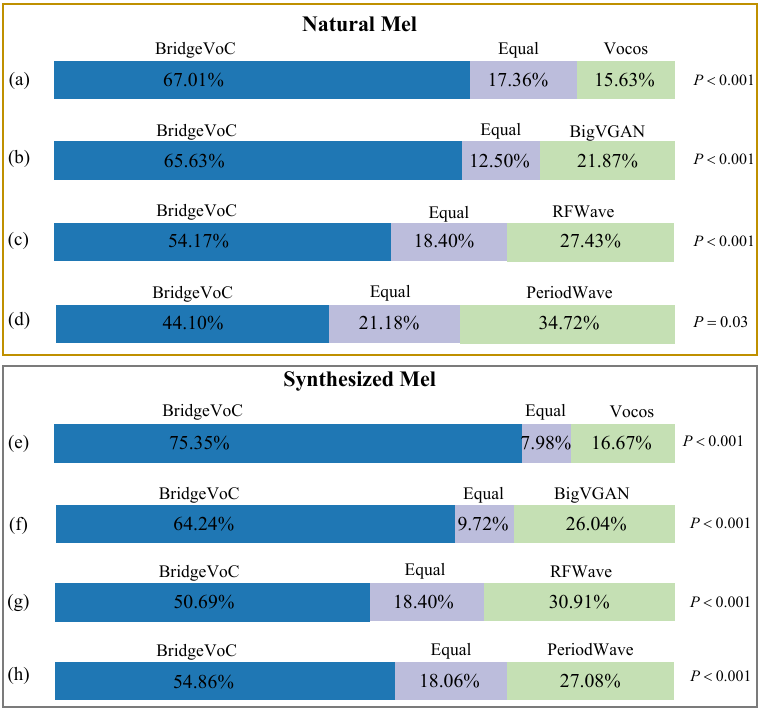}
	\vspace{-6pt}
	\caption{Average preference scores (in \%) between the proposed BridgeVoC and four baselines.}
	\label{fig:abtest}
	\vspace{-0.4cm}
\end{figure}
\renewcommand\arraystretch{0.85}
\begin{table*}[t]
	\centering
	\caption{Objective comparisons among different baselines on the LibriTTS benchmark. ``$a\times$'' denotes the speed-up ratio over real-time. Inference speed is evaluated on an Intel(R) Core(TM) i7-14700F and NVIDIA GeForce RTX 4060 Ti GPU. $\left(\cdot\right)^{\ddagger}$ indicates the use of pre-trained checkpoint from the original papers. $\left(\cdot\right)^{\star}$ denotes the single-step diffusion strategy is employed. The choice of NFE values for DDPM- and FM-based baselines follow the original literature.}
	\large
	\resizebox{0.98\textwidth}{!}{
	\begin{tabular}{ccIcccccIcccccccc}
		\toprule
		\multirow{2}*{\textbf{Models}} &\multirow{2}*{\textbf{Type}} &\multirow{2}*{\textbf{NFE}} &\textbf{\#Param.} &\textbf{\#MACs} &\multicolumn{2}{c|}{\textbf{Inference Speed}} &\multirow{2}*{M-STFT$\downarrow$} &\multirow{2}*{PESQ$\uparrow$} &\multirow{2}*{MCD$\downarrow$} &V/UV$\uparrow$ &Period.$\downarrow$ &\multirow{2}*{UTMOS$\uparrow$}  &\multirow{2}*{VISQOL$\uparrow$} &\multirow{2}*{WER$\downarrow$} \\
		\cmidrule{6-7}
		& & &\textbf{(M)} &\textbf{(Giga/5s)} &\textbf{CPU} &\textbf{GPU} & & & &F1 &RMSE &  & & \\
		\midrule
		GT &- &- &- &- &- &- &- &- &- &- &- &3.873 &- &6.21\\
		HiFiGAN-V1 &\multirow{8}*{GAN} &1 &14.01 &166.41 &5.88$\times$ &157$\times$ &1.104 &3.056 &4.284 &0.921 &0.167 &3.511 &4.721 &7.89 \\
		iSTFTNet-V1 &  &1 &13.33 &117.30 &10.66$\times$ &299$\times$ &1.157 &2.880 &4.480 &0.918 &0.167 &3.464 &4.655 &7.54 \\
		Avocodo &  &1 &14.01 &166.47 &5.29$\times$ &148$\times$ &1.113 &3.217 &4.314 &0.913 &0.161 &3.358 &4.762 &7.25\\
		BigVGAN-base &  &1 &14.01 &166.41 &2.33$\times$ &23$\times$ &0.884 &3.521 &3.034 &0.941 &0.131 &3.541 &4.869 &6.77 \\
		BigVGAN & &1 &112.39 &454.08 &1.54$\times$ &17$\times$ &0.812 &3.991 &2.547 &0.955 &0.104 &\underline{3.658} &4.934 &\underline{6.29}\\
		APNet & &1 &73.33 &33.92 &31.97$\times$ &499$\times$ &1.273 &2.897 &4.150 &0.927 &0.159 &2.416 &4.666 &7.54\\
		APNet2 & &1 &31.52 &\underline{14.79} &\underline{58.82$\times$} &\underline{749$\times$} &1.143 &2.834 &4.153 &0.923 &0.153 &2.956 &4.582 &7.44 \\
		Vocos$^{\ddagger}$ & &1 &13.53  &\textbf{6.35} &\textbf{142.76$\times$} &\textbf{1538$\times$} &0.854 &3.615 &3.105 &0.948 &0.115 &3.566 &4.879 &6.61\\
		\midrule
		FreGrad &\multirow{2}*{DDPM} &50 &\textbf{1.82}  &1589 &2.32$\times$ &35$\times$ &1.397 &3.793 &6.025 &0.929 &0.143 &3.217 &4.699 &8.26\\
		PriorGrad & &50 &\underline{2.70}  &8364 &0.56$\times$ &8$\times$ &1.393 &4.017 &5.854 &0.938 &0.130 &3.329 &4.737 &7.52\\
		\midrule
		PeriodWave$^{\ddagger}$ &\multirow{4}*{FM} &16 &29.81 &4986.56 &0.16$\times$ &4$\times$ &1.021 &4.240 &2.416 &0.959 &0.099 &3.652 &4.749 &6.66\\
		RFWave$^{\ddagger}$ & &10 &18.04  &598 &1.92$\times$ &26$\times$ &0.955 &4.251 &2.319 &0.961 &0.103 &3.362 &4.775 &6.50\\
		WaveFM$^{\ddagger}$ & &6 &19.50  &1188 &0.66$\times$ &13$\times$ &0.842 &3.954 &2.582 &0.955 &0.105 &3.191 &4.943 &6.48\\
		WaveFM$^{\star\ddagger}$ & &1 &19.50  &198.08 &4.24$\times$ &156$\times$ &0.883 &3.559 &2.931 &0.944 &0.134 &2.980 &4.895 &6.54\\
		\midrule
		\textbf{BridgeVoC(Ours)} &ScB &4 &7.65 &171.68 &3.28$\times$ &39$\times$ &\textbf{0.708} &\textbf{4.427} &\textbf{1.534} &\textbf{0.976} &\textbf{0.061} &\textbf{3.746} &\textbf{4.963} &6.64\\
		\midrule
		\textbf{BridgeVoC$^{\star}$(Ours)} &ScB &1 &7.65 &42.92 &13.91$\times$ &128$\times$ &0.738 &\underline{4.351} &\underline{1.717} &\underline{0.975} &\underline{0.066} &3.635 &\underline{4.957} &\textbf{6.10}\\
		\bottomrule
	\end{tabular}}
	\vspace{-0.3cm}
	\label{tbl:objective-metric-libritts}
\end{table*}
\renewcommand\arraystretch{0.9}
\begin{table}[t]
	\centering
	\caption{Objective comparisons among other baselines on the LJSpeech benchmark. Results for RFWave and WaveFM are not reported due to unavailable pre-trained checkpoints.}
	\huge
	\resizebox{\columnwidth}{!}{
	\begin{tabular}{ccIcccccc}
		\hline
		\multirow{2}*{\textbf{Models}} &\multirow{2}*{\textbf{NFE}}&\multirow{2}*{M-STFT$\downarrow$} &\multirow{2}*{PESQ$\uparrow$} &V/UV$\downarrow$ &\multirow{2}*{UTMOS$\uparrow$} &\multirow{2}*{VISQOL$\uparrow$} &\multirow{2}*{WER$\downarrow$}\\
		& & & &F1 & & &\\
		\hline
		GT &- &-  &- &- &4.376 &- &3.82\\
		HiFiGAN-V1 &1 &1.167  &3.574 &0.947 &4.218 &4.771 &4.22\\
		iSTFTNet-V1 &1 &1.188  &3.535 &0.947 &4.236 &4.756 &4.13 \\
		Avocodo$^{\ddagger}$ &1  &1.165 &3.604 &0.946 &4.152 &4.771 &4.01 \\
		BigVGAN-base &1  &0.978 &3.603 &0.956 &4.208 &4.822 &3.91\\
		BigVGAN  &1 &\underline{0.900} &4.107 &\underline{0.972} &\underline{4.314} &\underline{4.958} &\textbf{3.85}\\
		APNet &1   &1.266 &3.390 &0.945 &3.170 &4.695 &3.97\\
		APNet2 &1 &0.982 &3.492 &0.959 &3.984 &4.752 &4.15 \\
		Vocos &1  &1.012 &3.522 &0.956 &3.968 &4.774 &3.95 \\
		FreGrad &50 &1.344 &3.775 &0.943 &3.934 &4.574 &4.03\\
		PriorGrad  &50 &1.327 &3.955 &0.950 &4.006 &4.659 &4.08 \\
		PeriodWave$^{\ddagger}$ &16  &1.098 &\underline{4.235} &0.967 &\textbf{4.374} &4.722 &\underline{3.86} \\
		\textbf{BridgeVoC(Ours)} &4 &\textbf{0.888} &\textbf{4.385}  &\textbf{0.979} &4.302 &\textbf{4.963}
		&3.99\\
		\hline
	\end{tabular}}
	\vspace{-0.45cm}
	\label{tbl:objective-metric-ljs}
\end{table}
\renewcommand\arraystretch{0.85}
\begin{table*}[t]
	\centering
	\caption{Objective comparisons among different baselines on the out-of-distribution EARS and AISHELL3 datasets. The WER results on the EARS dataset are not reported due to unavailable transcripts.}
	\huge
	\resizebox{0.99\textwidth}{!}{
		\begin{tabular}{ccccccccIcccccccc}
			\toprule
			 \textbf{Test sets} &\multicolumn{7}{c}{\textbf{EARS (English with Diverse Emotions)}} &\multicolumn{8}{c}{\textbf{AISHELL3 (Mandarin)}}\\
			\midrule
			\multirow{2}*{\textbf{Models}}&\multirow{2}*{M-STFT$\downarrow$} &\multirow{2}*{PESQ$\uparrow$} &\multirow{2}*{MCD$\downarrow$} &V/UV$\uparrow$ &Period.$\downarrow$ &\multirow{2}*{UTMOS$\uparrow$}  &\multirow{2}*{VISQOL$\uparrow$} &\multirow{2}*{M-STFT$\downarrow$} &\multirow{2}*{PESQ$\uparrow$} &\multirow{2}*{MCD$\downarrow$} &V/UV$\uparrow$ &Period.$\downarrow$ &\multirow{2}*{UTMOS$\uparrow$}  &\multirow{2}*{VISQOL$\uparrow$} &\multirow{2}*{CER$\downarrow$} \\
			& & & &F1 &RMSE &  & & & & &F1 &RMSE &  & & \\
			\midrule
			GT &- &- &- &- &- &3.300 &- &- &- &- &- &- &2.700 &- &15.27\\
			HiFiGAN-V1 &1.210 &2.907 &2.733 &0.876 &0.149 &2.835 &4.570 &1.059 &2.823 &2.827 &0.934 &0.163 &2.432 &4.651 &15.59\\
			iSTFTNet-V1 &1.245 &2.604 &2.819 &0.851 &0.157 &2.771 &4.480 &1.132 &2.595 &3.061 &0.932 &0.164 &2.351 &4.486 &15.64\\
			Avocodo &1.197 &3.065 &2.703 &0.866 &0.149 &2.809 &4.641 &1.040 &3.003 &2.864 &0.937 &0.154 &2.126 &4.731 &15.59\\
			BigVGAN-base &1.025 &3.492 &1.871 &0.920 &0.106 &3.005 &4.739 &0.858 &3.414 &1.902 &0.958 &0.118 &2.411 &4.806 &15.37\\
			BigVGAN &\underline{0.839} &3.892 &1.400 &0.946 &\underline{0.083} &3.102 &4.898 &\underline{0.769} &3.878 &1.602 &\underline{0.963} &\underline{0.103} &\underline{2.520} &4.907 &15.41\\
			APNet2 &1.202 &2.450 &2.768 &0.862 &0.149 &2.259 &4.456 &0.975 &2.846 &2.743 &0.950 &0.136 &2.129 &4.715 &16.04\\
			Vocos$^{\ddagger}$ &0.898 &3.522 &1.949 &0.937 &0.094 &3.007 &4.848 &0.845 &3.342 &2.046 &0.952 &0.129 &2.409 &4.836 &15.41\\
			FreGrad &1.153 &3.495 &1.908 &0.869 &0.138 &2.753 &4.526 &1.061 &3.667 &1.930 &0.937 &0.148 &2.145 &4.621 &15.46\\
			PriorGrad &1.062 &3.847 &1.692 &0.849 &0.145 &2.863 &4.636 &1.054 &3.995 &1.717 &0.917 &0.160 &2.197 &4.675 &\textbf{15.28}\\
			PeriodWave$^{\ddagger}$ &0.971 &\underline{4.238} &1.171 &0.928 &0.103 &\textbf{3.218} &4.686 &0.958 &\underline{4.195} &1.598 &0.953 &0.119 &\textbf{2.565} &4.616 &\underline{15.32}\\
			RFWave$^{\ddagger}$ &0.879 &4.154 &\underline{1.143} &0.924 &0.112 &2.888 &4.750 &0.882 &4.145 &\underline{1.538} &0.950 &0.130 &2.050 &4.766 &15.46\\
			WaveFM$^{\ddagger}$ &0.850 &3.856 &1.553 &\underline{0.944} &0.091 &2.709 &\underline{4.926} &0.832 &3.810 &1.599 &0.964 &0.108 &2.037 &\underline{4.929} &15.37\\
			\midrule
			\textbf{BridgeVoC(Ours)} &\textbf{0.670} &\textbf{4.398} &\textbf{0.696} &\textbf{0.949} &\textbf{0.073} &\underline{3.159} &\textbf{4.948} &\textbf{0.649} &\textbf{4.382} &\textbf{0.950} &\textbf{0.966} &\textbf{0.092} &2.515 &\textbf{4.942} &15.46\\
			\bottomrule
	\end{tabular}}
	\vspace{-0.3cm}
	\label{tbl:objective-metric-ears-aishell3}
\end{table*}
\begin{table*}[t]
	\centering
	\caption{Subjective MOS among different vocoders on the out-of-distribution AISHELL3 and MUSDB18 benchmarks. We report mean with 95\% confidence intervals (CI95), and we perform a t-test comparing BridgeVoC and RFWave.}
	\resizebox{0.8\textwidth}{!}{
	\begin{tabular}{cccccccc}
		\hline
		Models &GT &Vocos &BigVGAN &WaveFM &PeriodWave &RFWave &BridgeVoC(Ours) \\
		AISHELL3 &4.43$\pm$0.05 &3.96$\pm$0.08 &4.02$\pm$0.07 &3.64$\pm$0.05 &4.09$\pm$0.07 &4.12$\pm$0.08 &\textbf{**4.24$\pm$0.04}\\
		MUSDB18(Vocals) &4.39$\pm$0.04 &3.89$\pm$0.07    &4.00$\pm$0.06 &3.58$\pm$0.05 &4.08$\pm$0.06 &4.13$\pm$0.06 &\textbf{**4.20$\pm$0.05} \\
		\hline \\[-0.8em] \multicolumn{8}{l}{Note: $^{**}p<0.05$, $^{*}p<0.1$}\\
	\end{tabular}}
	\label{tbl:libritts-mos}
	\vspace{-0.45cm}
\end{table*}
\begin{figure*}[t]
	\centering
	\vspace{0pt}
	\includegraphics[width=0.8\textwidth]{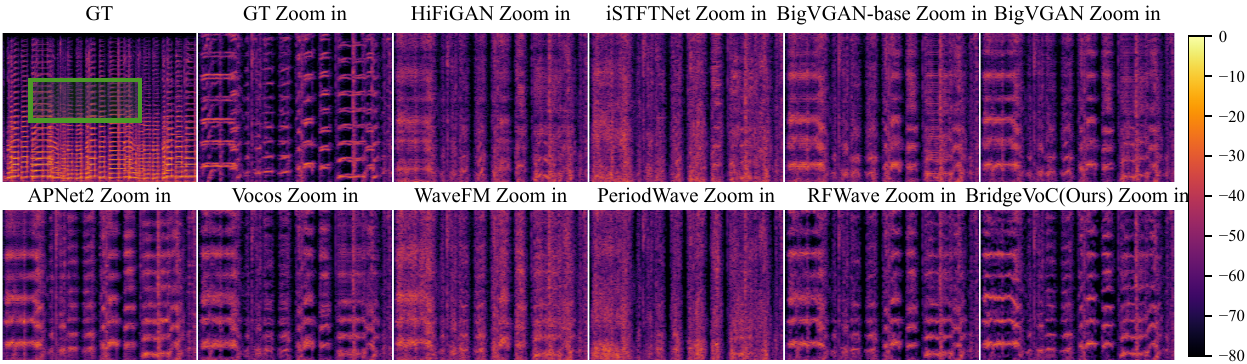}
	\vspace{-2pt}
	\caption{Spectral visualization of different vocoder methods on an out-of-distribution musical sound from the FSD50K dataset.}
	\label{fig:sota-comparison-effect-visualization}
	\vspace{-10pt}
\end{figure*}
\begin{figure*}[t]
	\centering
	\vspace{0pt}
	\includegraphics[width=0.8\textwidth]{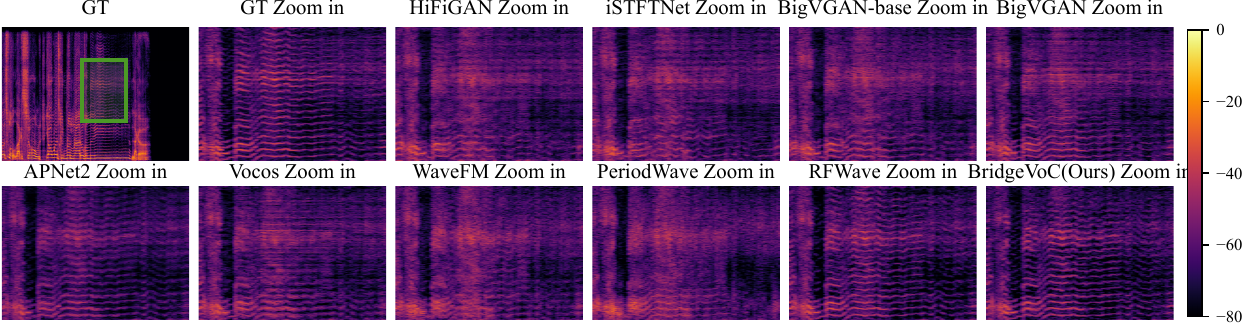}
	\vspace{-2pt}
	\caption{Spectral visualization of different vocoder methods on an out-of-distribution vocal sound from the MUSDB18 dataset.}
	\label{fig:sota-comparison-vocals-visualization}
	\vspace{-10pt}
\end{figure*}
\begin{figure*}[t]
	\centering
	\vspace{0pt}
	\includegraphics[width=0.99\textwidth]{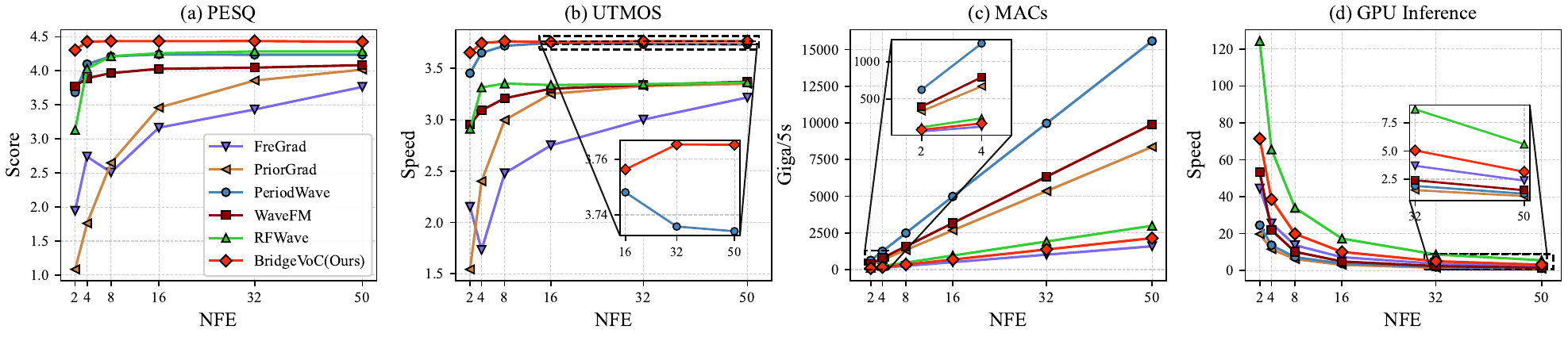}
	\vspace{-2pt}
	\caption{Reconstruction quality and inference efficiency between BridgeVoC and another diffusion baselines under different NFE setups.}
	\label{fig:metrics-comparisons}
	\vspace{-0.3cm}
\end{figure*}
\begin{figure*}[t]
	\centering
	\vspace{0pt}
	\includegraphics[width=0.88\textwidth]{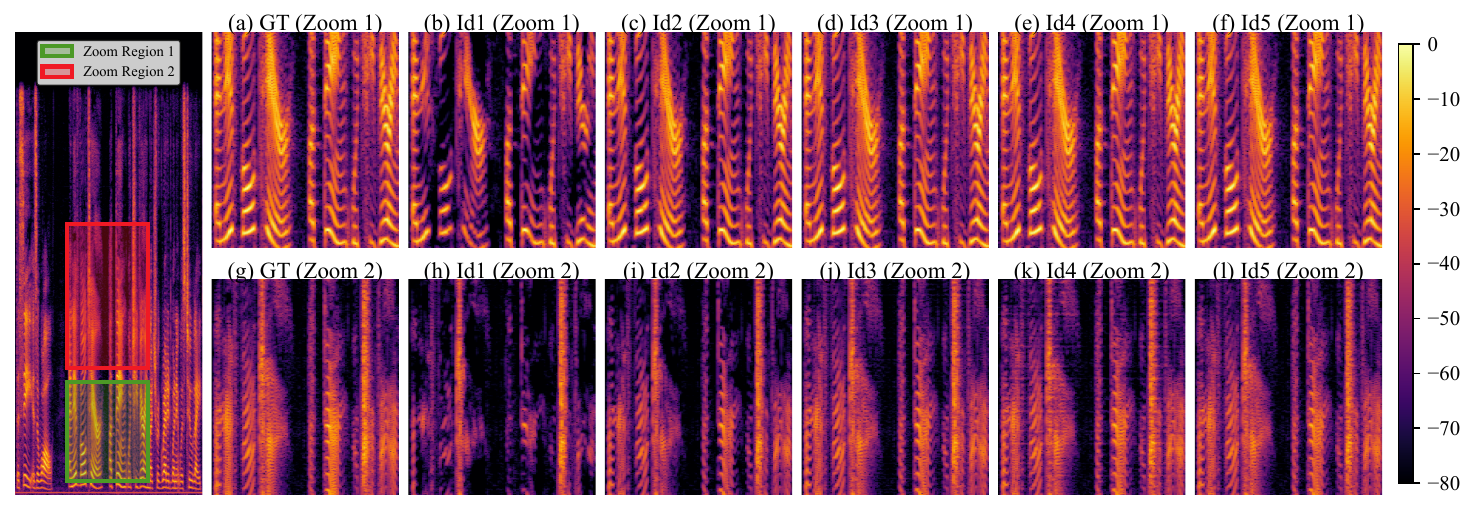}
	\vspace{-7pt}
	\caption{Spectral visualization \emph{w.r.t.} different loss setups for single-step-distillation. The audio clip is from the LibriTTS benchmark. (a) Zoom-in spectrum of the target for region 1. (b)-(f) Zoom-in spectra from Id1 to Id5 for region 1. (g) Zoom-in spectrum of the target for region 2. (h)-(l) Zoom-in spectra from Id1 to Id5 for region 2.}
	\label{fig:single_step_visualization}
	\vspace{-0.2cm}
\end{figure*}
\vspace{-8pt}
\section{Experimental Results and Analysis}\label{sec:experiments-analysis}
\vspace{-3pt}
\subsection{Ablation Experiments}\label{sec:ablations}
\textbf{Different noise schedulers, losses, and inference samplers:} Table~{\ref{tab:ablations}} reports the objective evaluation scores under different noise schedulers, loss setups and inference samplers. We first explore the influence of loss configurations, where the noise scheduler and sampler are fixed to gmax and SDE, respectively. When only the data prediction loss $\mathcal{L}_{data}$ is used, the worst performance is observed across all metrics, with significant information loss of spectral details, as shown in Fig.~{\ref{fig:ablation_loss_visualization}}(b). 
From Id2 to Id3, incorporating the Mel loss $\mathcal{L}_{mel}$ drastically facilitates 
spectral reconstruction, with objective scores like PESQ and UTMOS improving noticeably. However, we observe prominent point-like artifacts in the mid- and high-frequency regions of the generated spectra, as shown in Fig.~{\ref{fig:ablation_loss_visualization}}(c). To mitigate that, we incorporate the adversarial loss to further suppress the artifacts and improve the generation quality, as shown in Fig.~{\ref{fig:ablation_loss_visualization}}(d). We then compare three representative noise schedulers: gmax, VE and VP, paired with two inference samplers (SDE and ODE).
For the SDE sampler, the gmax scheduler outperforms VE and VP across all metrics, while the VE scheduler becomes optimal when paired with the ODE sampler. Notably, the VP-ODE combination (Id8) is excluded from results, as it fails to generate audible audio. Based on these findings, we choose the configuration of Id3 as the default for subsequent experiments.

\textbf{Different Network Architectures:} In Table~{\ref{tbl:ablations-network}}, we compare the 
performance of two candidate networks, \emph{i.e.}, NCSN++ and BCD, alongside with the number of trainable parameters and multiply-accumulate operations (MACs). Metric results are based on the test set of the LibriTTS benchmark. Both architectures deliver competitive performance, and BCD achieves a superior efficiency-quality trade-off. For $\text{NFE}=4$, BCD uses only 4.38\% of the MACs of the larger NCSN++ variant (64.8 M) while outperforming it in multiple metrics, confirming the effectiveness of BCD in subband prior modeling. Unless otherwise specified, the proposed BCD is adopted as the default network architecture in subsequent experiments.

We further validate the core component of BCD, the LKCAB, by comparing it to two alternative blocks with consistent computational cost: Set-A replaces the LKCAB with ConvNeXt blocks, and Set-B replaces the LKCAB's large-kernel convolution with local self-attention~{\cite{liu2021swin}}. Table~{\ref{tbl:ablations-basic-architecture}} shows that the proposed LKCAB outperforms both variants, achieving lower M-STFT and higher V/UV F1, demonstrating the modeling advantage of our LKCAB. We also present the results under different  $\left\{k_{f}, k_{l}\right\}$ configurations. As shown in Fig.~{\ref{fig:ablation_kernel_size}}, a suitable receptive field is significant for spectrum restoration. While small kernels, \emph{e.g.}, (3, 5), fail to model inter-frame and cross-subband dependencies, excessively large kernels, \emph{e.g.}, (15, 17), may overlook local fine-grained details like high-frequency harmonics. This receptive trade-off strikes a performance peak as the kernel size gradually increases, and $\left\{k_{f}=9, k_{l}=11\right\}$ is selected by default due to its empirical optimal performance.

\textbf{Different number of divided subbands:} Given that subband modeling is central to the processing efficiency of BCD, we explore  the effect of different division strategies. We adjust the kernel/stride of the Conv2d in Eq.~{\ref{eqn:22}} and Transposed Conv2d in Eq.~{\ref{eqn:26}} to yield varying number of subbands $N$, thereby controlling the subband division granularity. Additionally, we compare these results with an even spectrum division tactic. As shown in Table~{\ref{tbl:objective-different-subband}}, increasing $N$ notably improves quality, as more subbands enable finer frequency modeling, reducing spectral smearing and enhancing harmonic reconstruction. As shown in Fig.~{\ref{fig:ablation_nbands_visualization}}, mid-frequency harmonic components become more progressively clear as $N$ increases, but also at the cost of increasing processing cost. As a trade-off between quality and inference cost, we adopt 24 subbands by default. Besides, for the same number of subbands, the even division strategy leads to marked performance degradation, validating the superiority of the proposed uneven splitting method. It should be noted that, unlike traditional parameter scaling~{\cite{leebigvgan}}, we boost the performance by scaling up the number of subbands, and thus we term the scheme as \textbf{subband scaling}, which has been rarely investigated in the audio generation task.  
\renewcommand\arraystretch{0.85}
\begin{table*}[t]
	\centering
	\caption{The ablation studies in terms of the adopted loss for single-step distillation. The results are based on the test set of LibriTTS benchmark.}
	\label{tbl:single-step-ablas}
	\LARGE
	\resizebox{0.95\textwidth}{!}{
		\begin{tabular}{cIcccccIcccccccc}
			\toprule
			\multirow{2}*{\textbf{Ids}}   &\multirow{2}*{$\boldsymbol{\mathcal{L}_{naivedistill}}$} &\multirow{2}*{$\boldsymbol{\mathcal{L}_{omnidistill}}$} &\multirow{2}*{$\boldsymbol{\mathcal{L}_{mel}}$}
			&\multirow{2}*{$\boldsymbol{\mathcal{L}_{g}+\mathcal{L}_{fm}}$}
			&\multirow{2}*{$\boldsymbol{\mathcal{L}_{inverse}+\mathcal{L}_{gt}}$}
			&\multirow{2}*{M-STFT$\downarrow$} &\multirow{2}*{PESQ$\uparrow$} &\multirow{2}*{MCD$\downarrow$}  &Periodicity$\downarrow$ &\multirow{2}*{V/UV F1$\uparrow$} &\multirow{2}*{UTMOS$\uparrow$} &\multirow{2}*{VISQOL$\uparrow$} &\multirow{2}*{WER$\downarrow$}\\
			& & & & & & & & &RMSE & & & &\\
			\midrule
			1 &\Checkmark &\XSolidBrush &\XSolidBrush &\XSolidBrush &\XSolidBrush &1.468 &2.614 &4.590 &0.222 &0.881 &3.316 &4.067 &9.38\\
			2 &\XSolidBrush &\Checkmark &\XSolidBrush &\XSolidBrush &\XSolidBrush &1.064 &4.182 &2.412 &0.130 &0.946 &3.641 &4.678 &6.88\\
			3 &\XSolidBrush &\Checkmark &\Checkmark &\XSolidBrush &\XSolidBrush &\textbf{0.714} &\textbf{4.363} &\underline{1.669} &\textbf{0.063} &\underline{0.975} &\underline{3.650} &\textbf{4.963} &\underline{6.26}\\
			4 &\XSolidBrush &\Checkmark &\Checkmark &\Checkmark &\XSolidBrush &0.731 &\underline{4.355} &\textbf{1.665} &\underline{0.064} &\textbf{0.976} &\textbf{3.668} &\underline{4.959} &6.45\\
			5 &\XSolidBrush &\Checkmark &\Checkmark &\Checkmark &\Checkmark &0.738 &4.351 &1.717 &0.066 &0.975 &3.635 &4.957 &\textbf{6.10}\\
			\bottomrule
	\end{tabular}}
	\vspace{-4pt}
\end{table*}
\vspace{-12pt}
\subsection{Compared with State-of-the-art Methods}\label{sec:comparisons-sota}
\vspace{-2pt}
This section presents a comprehensive analysis of the proposed BridgeVoC, covering both objective and subjective evaluations across multiple datasets to validate its superiority. 

\textbf{Objective Evaluations:} Table~{\ref{tbl:objective-metric-libritts}} presents performance on the LibriTTS benchmark, alongside with network parameters, computational complexity, and inference speed. A key observation is that diffusion-based methods (DDPM and FM) generally achieve better reconstruction quality than GAN-based counterparts. For example, very recent FM-based models like PeriodWave and RFWave outperform GAN-based HiFiGAN and Vocos by a large margin, and also surpass BigVGAN in several metrics.  Nonetheless, this quality advantage often comes with much higher computational costs and slower inference speeds. BridgeVoC addresses it by integrating subband modeling and refined spectral processing, that is, ScB narrows the generation trajectory with fewer diffusion steps, and subband modeling enables fine-grained spectral reconstruction with lower computational overhead, thus resulting in both high quality and lightweight design. Specifically, with only 4 function evaluations, BridgeVoC achieves SoTA performance across most metrics, and the only except is WER, revealing an inherent inconsistency between acoustic reconstruction accuracy and semantic information preservation. Table~{\ref{tbl:objective-metric-ljs}} summarizes performance on the LJSpeech dataset, and similar superiority is observed, validating the effectiveness across datasets.

To evaluate generalization capability, pre-trained models on the LibriTTS benchmark are tested on 
out-of-distribution EARS and AISHELL3 test sets. Table~{\ref{tbl:objective-metric-ears-aishell3}} reports the performance. Again, BridgeVoC maintains notable superiority in domain generalization, outperforming all baselines by a large margin in metrics like M-STFT and PESQ, while achieving suboptimal performance in UTMOS and CER. Besides, we notice relatively large score gap between the two datasets in UTMOS, revealing the limitation of UTMOS in cross-lingual generalization.

\textbf{Subjective Evaluations:} We conduct similarity MOS (SMOS) evaluations from two out-of-distribution sets: AISHELL-3 and MUSDB18, and 12 clips are sampled for each set. For MUSDB18, we choose the vocal type as it includes rich harmonic structures and can amplify the difference between different algorithms. Table~{\ref{tbl:libritts-mos}} shows that BridgeVoC achieves consistent superiority over both GAN- and FM-based methods on both sets, confirming its advantage in human-perceived quality. A notable finding is that, despite WaveFM exhibits promising objective results, it yields significantly lower MOS scores. According to the feedback from participants, audible high-frequency current noise can exist in its generated audio. This discrepancy underscores the inherent gap between objective metrics and subjective perception. 

We present the AB testing results between our method and the other four baselines. For the natural scenario, Mel spectra are extracted from the MUSDB18 vocal clips, and for the synthetic case, Mels are generated via the music generation model~{\cite{Novack2024Ditto}}. As shown in Fig.~{\ref{fig:abtest}}, for both cases, our method enjoys consistent and significant preference over baselines, validating its robustness to different input types. The details of the subjective evaluation platform are provided in the supplemental materials.

\textbf{Spectral Visualizations:} Figs.~{\ref{fig:sota-comparison-effect-visualization}} to {\ref{fig:sota-comparison-vocals-visualization}} present qualitative comparisons of different vocoders, which are pre-trained on the LibriTTS benchmark. For musical sounds, \emph{e.g.}, aerophones, our method outperforms other approaches by better reconstructing harmonic components (Fig.~{\ref{fig:sota-comparison-effect-visualization}}). For vocal sounds (Fig.~{\ref{fig:sota-comparison-vocals-visualization}}), while other methods exhibit varying degrees of distortion in harmonic-rich regions, our method can effectively recover fine-grained details, as evident in the zoomed-in regions of the figures.

\textbf{Computational Efficiency:} Fig.~\ref{fig:metrics-comparisons} presents a comprehensive comparison between BridgeVoC and five advanced diffusion-based baselines regarding reconstruction quality and processing efficiency across varying number of function evaluations. For metrics including PESQ, UTMOS and inference speed, higher values indicate better performance, while lower values are preferred for MACs. A key observation from the figure is that BridgeVoC achieves promising generation quality even with extremely few inference steps, \emph{i.e.}, $\text{NFE}=2$, it already delivers PESQ and UTMOS scores that surpasses baselines requiring more steps, \emph{e.g.}, $\text{NFE}\geq 16$. As NFE further increases, BridgeVoC continues to converge stably in quality. Moreover, owing to the carefully designed subband architecture, our method requires substantially fewer MACs and yields faster inference, exhibiting prominent superiority in the trade-off between performance and inference cost.
\begin{figure}[t]
	\centering
	\vspace{0pt}
	\includegraphics[width=\columnwidth]{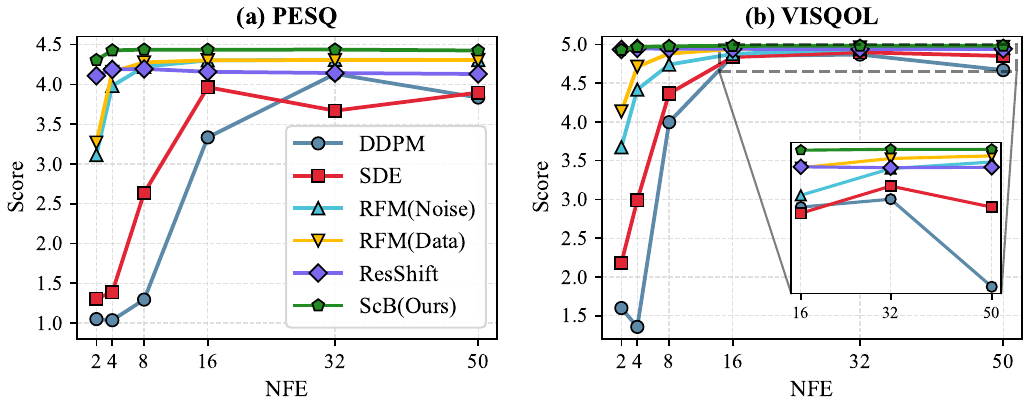}
	\vspace{-8pt}
	\caption{Generation performance between BridgeVoC and another diffusion types under different NFE setups. ``RFM(Noise)'' and ``RFM(Data)'' denote the N2D and D2D paradigm are adopted for RFM, respectively.}
	\label{fig:diffusion-comparisons}
	\vspace{-0.4cm}
\end{figure}
\begin{figure}[t]
	\centering
	\vspace{0pt}
	\includegraphics[width=\columnwidth]{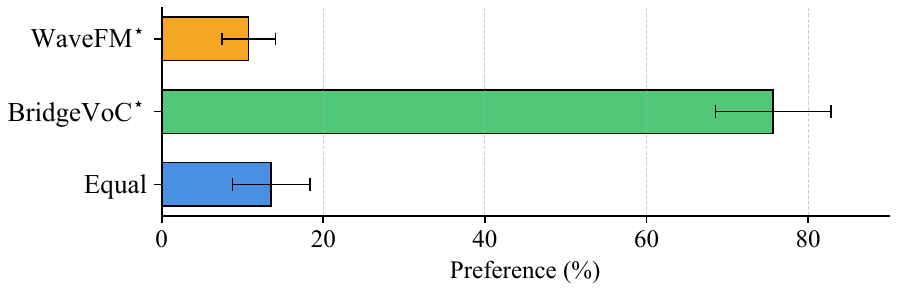}
	\vspace{-6pt}
	\caption{Average preference scores (in \%) between the proposed BridgeVoC with single-step distillation and WaveFM with consistency distillation algorithm.}
	\label{fig:single_step_abtest}
	\vspace{-0.5cm}
\end{figure}
\subsection{Performance of Different Diffusion Schemes}\label{sec:performance-different-diffusion}
\vspace{-2pt}
To validate the superiority brought by Schr\"odinger bridge model, we compare it with another four diffusion schemes: DDPM, SDE, Rectified FM (RFM), and ResShift~{\cite{yue2023resshift}}. For DDPM, the source is a standard Gaussian noise, \emph{i.e.}, N2D generation paradigm. For RFM, we consider two cases: N2D and D2D paradigms, where the latter is enabled by replacing the mean of the source distribution with the RSS component of the Mel-spectrum, say, $\mathbf{Y}=\mathcal{A}^{\dagger}\mathbf{Z}\exp\left(\mathbf{\Phi}_{init}\right)$. For SDE and ResShift, we adopt the D2D paradigm. Objective comparisons under varying NFEs are presented in Fig.~{\ref{fig:diffusion-comparisons}}. Several observations can be made. First, compared with DDPM and SDE, flow-matching, ResShift and bridge models yield better performance in generation quality and convergence. Second, comparing between the two paradigms of RFM, D2D paradigm exhibits relatively better generation quality, and the advantage seems more prominent when fewer NFE is adopted, indicating that the proposed D2D paradigm can notably narrow the generation trajectory and also improve the performance upper-bound. Besides, the Schr\"odinger Bridge shows the best performance, validating the intrinsic advantage of the bridge model in the vocoder task.
\vspace{-10pt}
\subsection{Performance on Single-Step Distillation}\label{sec:performance-single-step}
\vspace{-2pt}
Table~{\ref{tbl:single-step-ablas}} analyzes the performance impact of different loss configurations adopted in Sec.~{\ref{sec:distilled-single-step}}. Several valuable conclusions can be drawn. First, the proposed omnidirectional distillation improves the distillation performance by a large margin, as observed from Id1 to Id2. This is because the proposed omnidirectional differential operation alleviates the distillation difficulty incurred by the phase wrapping effect, thus facilitating the information transfer from the teacher to student model. As shown in Fig.~{\ref{fig:single_step_visualization}}(b)-(c) and (h)-(i), the adoption of the omnidirectional distillation notably recovers the distorted harmonic structures. Second, by incorporating the Mel loss, spectral details can be further complemented, as shown from Id2 to Id3 and Fig.~{\ref{fig:single_step_visualization}}(i)-(j). In addition, we empirically observe that the introduction of adversarial loss can boost the textual detail recovery in the high-frequency regions, but also cause the artifact in the mid-frequency regions (see Fig.~{\ref{fig:single_step_visualization}}(k)). And the introduction of the consistency-preservation loss can effectively suppress it, as shown in Fig.~{\ref{fig:single_step_visualization}}(l). Besides, as shown in Id5, the introduction of the consistency-preservation loss results in the lowest WER, indicating that this loss can effectively facilitate the recovery of the semantic information by means of the bijective mapping mechanism.

As shown in Table~{\ref{tbl:objective-metric-libritts}}, our single-step diffusion model, \emph{i.e.}, BridgeVoC$^{\star}$, achieves comparable performance over our multi-step version, and still outshines other baselines by a large margin, including WaveFM$^{\star}$, the single-step version of WaveFM{\footnote{To our best knowledge, WaveFM is the only open-sourced diffusion-based single-step neural vocoder to be compared.}}. For further comparison with WaveFM$^{\star}$, in Fig.~{\ref{fig:single_step_abtest}}, we conduct the AB testing between BridgeVoC$^{\star}$ and WaveFM$^{\star}$. The clips include both speech and music types. Evidently, our proposed method notably outperforms WaveFM$^{\star}$. Collectively, via the proposed single yet effective distillation strategy, our single-step diffusion models achieve promising performance in both objective and subjective evaluations. 
\vspace{-0.2cm} 
\section{Concluding Remarks}
\label{sec:conclusion}
This paper presents BridgeVoC, a novel time-frequency (T-F) diffusion vocoder. Specifically, we revisit the vocoder task through rank analysis, and formulate it as a specialized type of audio restoration task given the range-space spectral representation as the degraded input. Building on that, the Schr\"odinger Bridge is introduced, where the range-space spectral surrogate and target spectra serve as dual endpoints for stochastic trajectory modeling. To fully exploit the hierarchical prior of audio subbands in the T-F domain, we devise a subband-aware network dubbed BCD for data prediction. In this network, we adopt uneven strategy to partition spectral subbands and employ a convolutional-style attention module with large kernels, enabling efficient T-F contextual modeling. For single-step inference, an omnidirectional distillation loss is proposed to enable effective teacher-student knowledge transfer, and the performance is improved by incorporating target-related and bijective consistency losses. We conduct extensive experiments on the LJSpeech and LibriTTS benchmarks, as well as on out-of-distribution datasets. Both quantitative and qualitative experiments show that with only 4 sampling steps, the proposed vocoder substantially surpasses existing GAN-, DDPM- and flow-matching-based vocoders. The superiority is also achieved by the proposed single-step inference variant.\\
\textbf{Limitations:} While this paper focuses on audio vocoder from the Mel-spectrum, many recent literature also concentrate on the nonlinear latent features from variational auto-encoders (VAEs) or discrete codes from the neural audio codecs (NACs). Due to the linear degradation property of Mel-spectrum, we can elegantly obtain the RSS via the linear operation, which unfortunately does not hold in nonlinear scenarios. Therefore, further investigations are still required to expand the proposed vocoder framework into more general acoustic representations.

%
{
\bibliographystyle{IEEEtran}
\small
\bibliography{IEEEabrv, refs.bib}
}
\vspace{-10mm}
\begin{IEEEbiography}[{\includegraphics[width=1in,height=1.15in,clip,keepaspectratio]{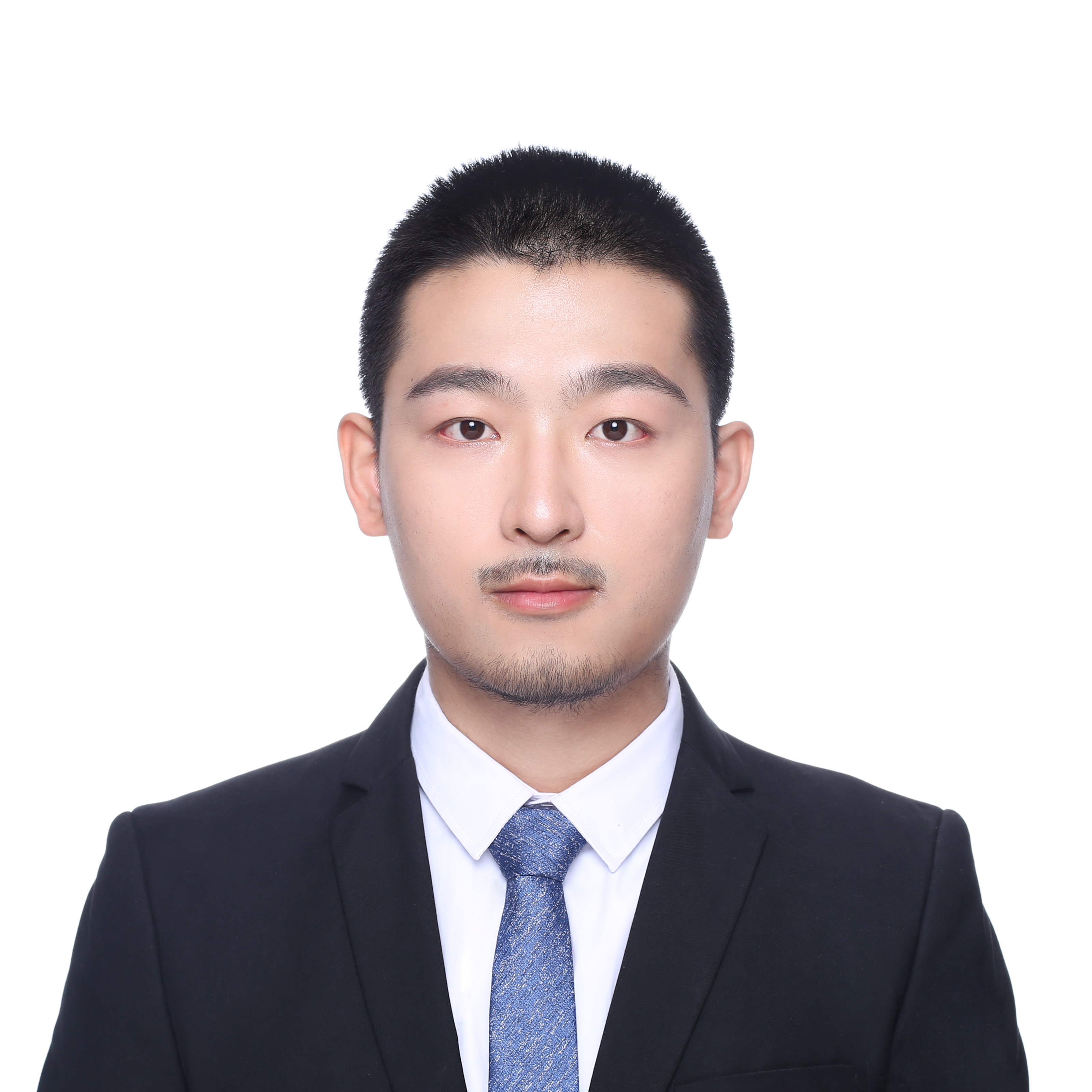}}]
	{Andong Li} 
	(Member, IEEE) received the B.S. degree in information engineering from Southeast University, Nanjing, China, in 2018, and the Ph.D. degree in signal and information processing from the Institute of Acoustics, Chinese Academy of Sciences, Beijing, China, in 2023. From 2023 to 2024, he was a Senior Researcher with Tencent AI Lab. He is currently an Associate Researcher with the Institute of Acoustics, Chinese Academy of Sciences. His research interests include speech enhancement, audio coding, array signal processing, and speech editing. He is also an active reviewer for multiple leading conferences and journals, such as INTERSPEECH, ICASSP, AAAI, ICLR, IJCAI, ACM MM, IEEE Signal Processing Letters, Speech Communication, Neural Networks, Pattern Recognition, and IEEE/ACM Transactions on Audio, Speech, and Language Processing.
\end{IEEEbiography}
\begin{IEEEbiography}[{\includegraphics[width=1in,height=1.15in,clip,keepaspectratio]{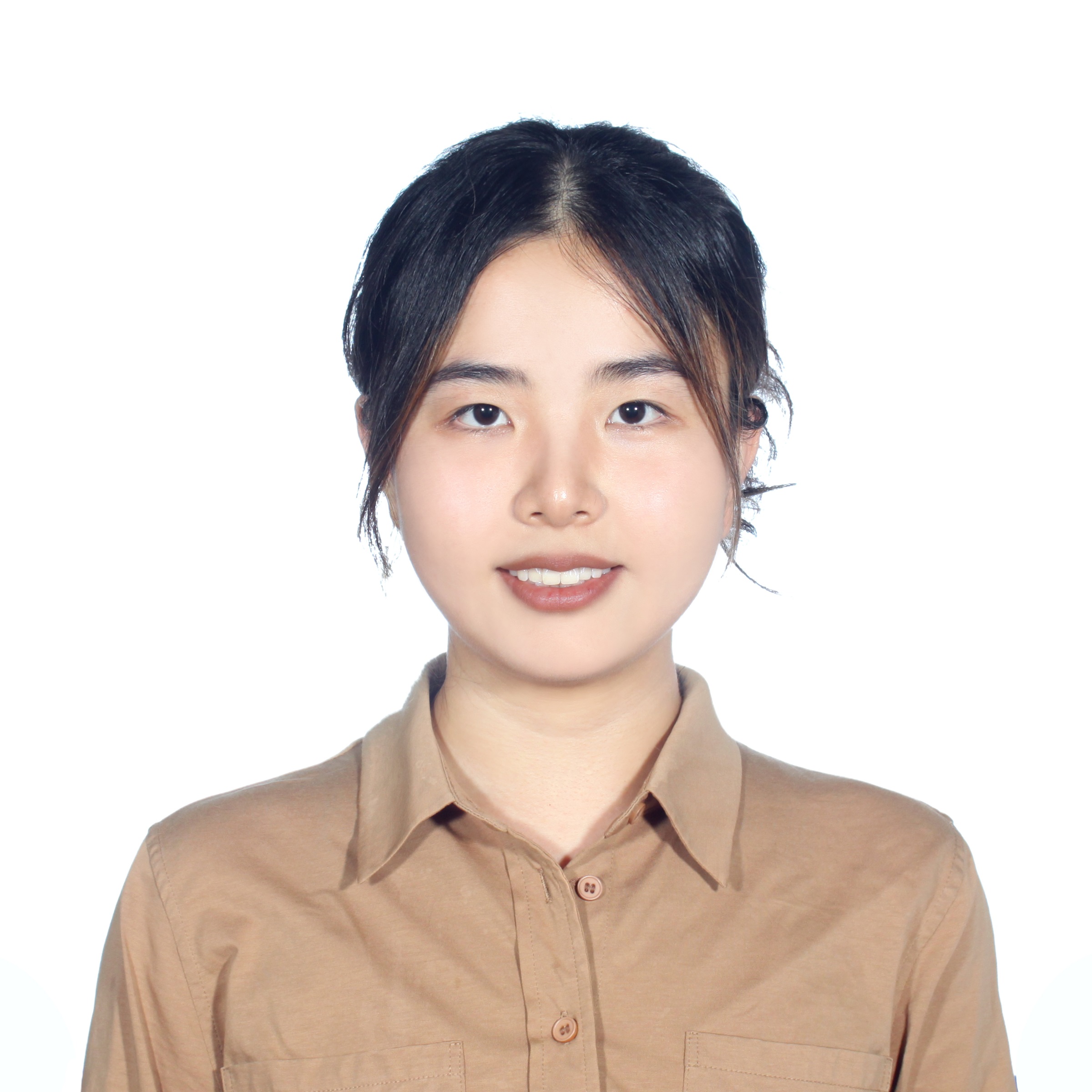}}]
	{Tong Lei} 
	Tong Lei received the B.Sc. degree in Physics from Nanjing University, Nanjing, China, in 2020. She received the Ph.D. degree from the Key Laboratory of Modern Acoustics, Nanjing University, in 2025. She joined Tencent AI Lab as a full‑time researcher in July 2025. Her research interests include speech enhancement, microphone‑array signal processing, audio understanding, and diffusion models.
\end{IEEEbiography}
\begin{IEEEbiography}[{\includegraphics[width=1in,height=1.15in,clip,keepaspectratio]{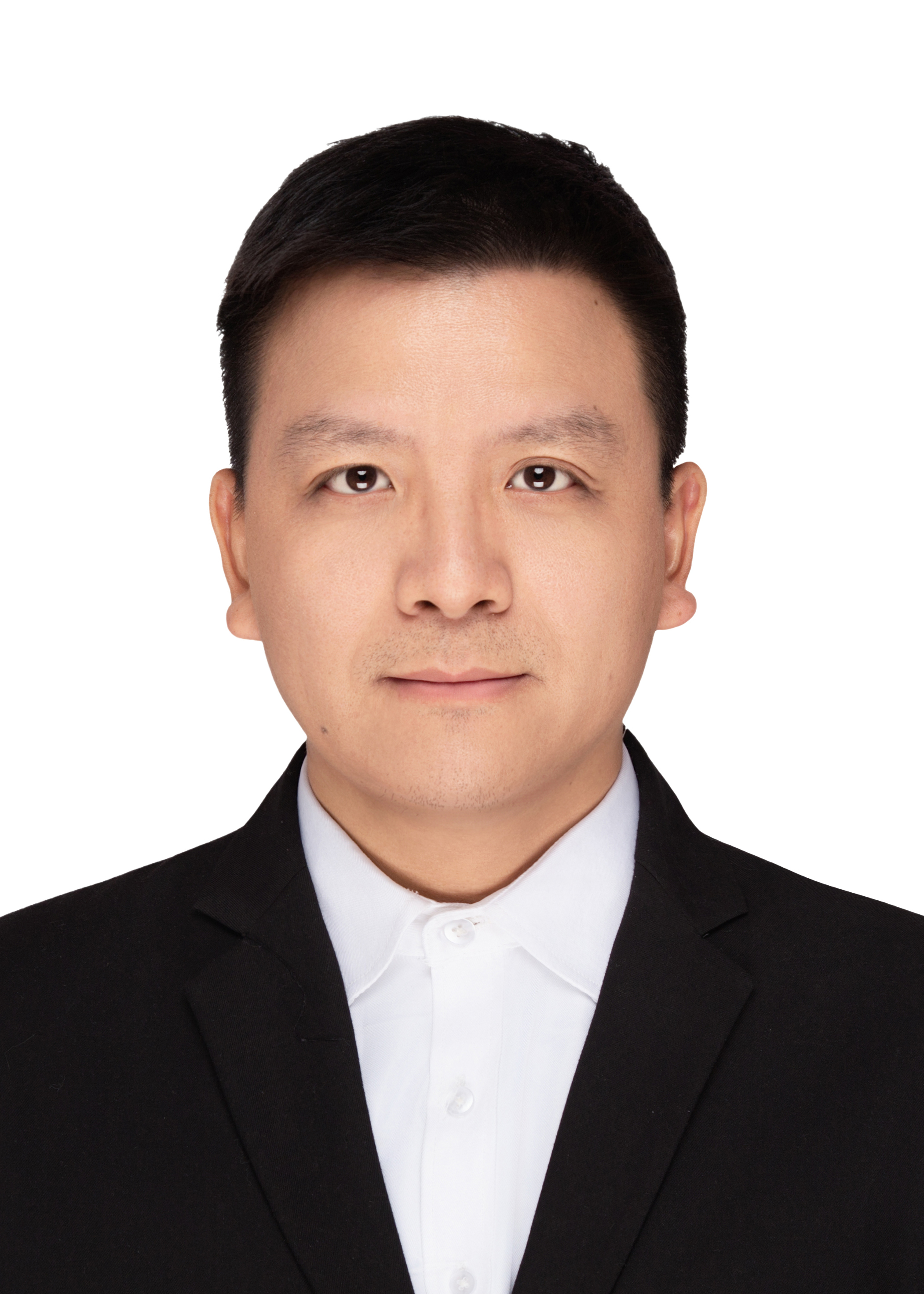}}]
	{Rilin Chen}
	received his B.Sc. degree in Communication Engineering from the School of Information Engineering in 2007, and later earned his Ph.D. in Acoustics from the Institute of Acoustics, Chinese Academy of Sciences in 2012. He is currently employed at Tencent Technology Co., Ltd. His research interests include speech enhancement, audio separation, and microphone array signal processing.
\end{IEEEbiography}
\begin{IEEEbiography}[{\includegraphics[width=1in,height=1.15in,clip,keepaspectratio]{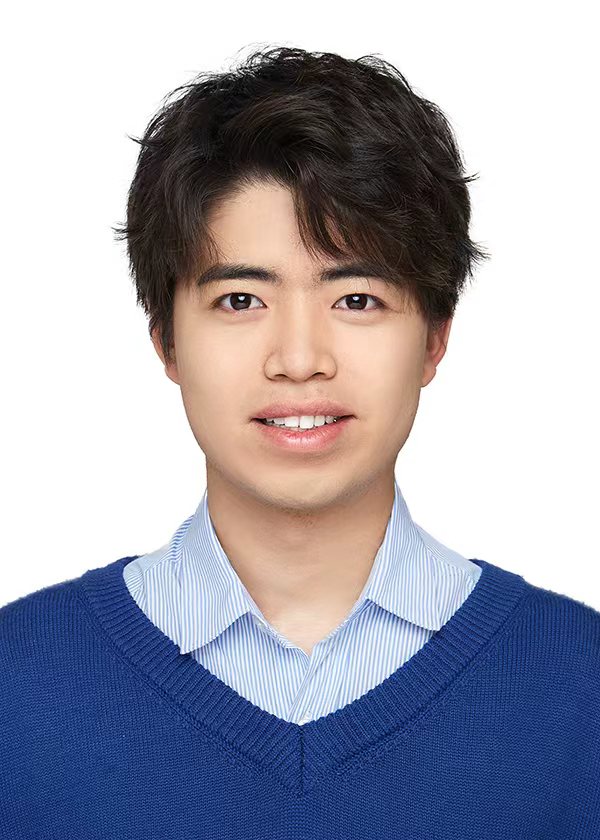}}]
	{Kai Li}
	(Student Member, IEEE) received the B.S. degree from the Department of Computer Technology and Application, Qinghai University, Xining, China, in 2020, and the M.S. degree from the Department of Computer Science and Technology, Tsinghua University, Beijing, China, in 2024, where he is currently pursuing the Ph.D. degree under the supervision of Prof. Xiaolin Hu. His current research interests include speech/music separation, multi-modal speech separation, and audio large language models. 
\end{IEEEbiography}
\begin{IEEEbiography}[{\includegraphics[width=1in,height=1.15in,clip,keepaspectratio]{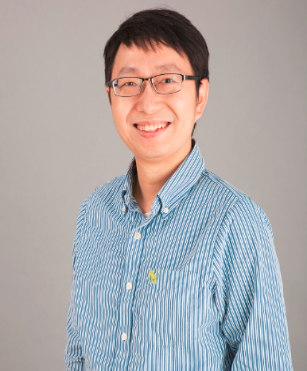}}]
{Meng Yu} became a Member of IEEE in 2023. He has been a Principal   Research Scientist at Tencent AI Lab since 2016. From 2013 to 2016, he 
worked as a Staff Research Engineer at Audience (a Knowles Company), focusing on audio/speech enhancement for voice communication and 
improving speech recognition. Prior to that, he was a Software Engineer at 
Cisco from 2012 to 2013, specializing in speaker segmentation and 
recognition. He received B.S. in Mathematics from Peking University, 
Beijing, China in 2007, and a Ph.D. degree in Mathematics from University of California, Irvine, CA, USA in 2012. His research interests focus on audio and speech processing, with a particular emphasis on single and multi-channel speech enhancement, dereverberation, echo cancellation, howling suppression, and far-field frontend speech enhancement applications. 
\end{IEEEbiography}
\begin{IEEEbiography}[{\includegraphics[width=1in,height=1.15in,clip,keepaspectratio]{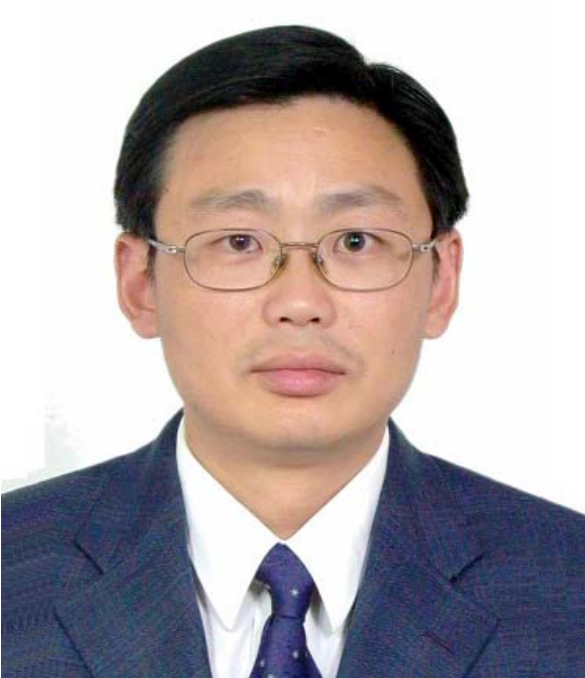}}]
	{Xiaodong Li}
	received the B.S. degree from Nanjing University, Nanjing, China, in 1988, the M.Eng. degree from the Harbin Engineering University, Harbin, China, in 1991, and the Ph.D. degree in physical acoustics from the Institute of Acoustics, Chinese Academy of Sciences (IACAS), Beijing, China, in 1995. After a short period as a Research Fellow with the Hong Kong Polytechnic University, Hong Kong, working on active control of sound, he was appointed Assistant Professor with IACAS in 1997. He was made an Associate Professor with IACAS in 1998 and Professor in 2002. His research interests include acoustic signal processing, active control of sound and vibration, and engineering acoustics.
\end{IEEEbiography}
\begin{IEEEbiography}[{\includegraphics[width=1in,height=1.15in,clip,keepaspectratio]{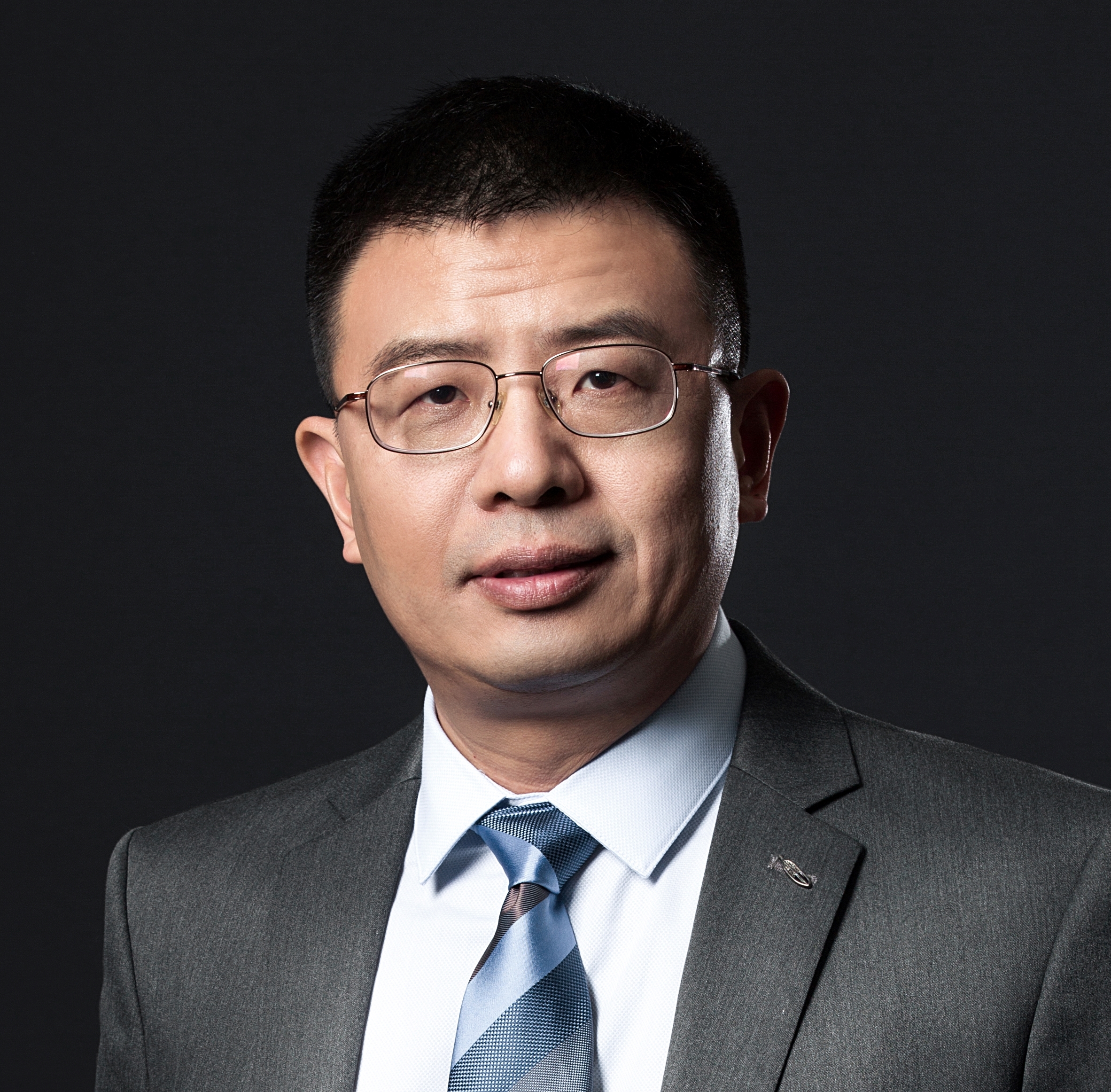}}]
	{Dong Yu}
	(Fellow, IEEE)
	is a Fellow of ACM, IEEE, and ISCA. He is currently a distinguished scientist and vice general manager at Tencent AI Lab. Prior to joining Tencent in 2017, he was a principal researcher at Microsoft Research (Redmond). Dr. Dong Yu’s research focuses on speech and natural language processing. He has published two monographs and over 350 papers. His works have been widely cited and recognized by the IEEE Signal Processing Society best transaction paper award in 2013, 2016, 2020, and 2022, the 2021 NAACL best long paper award, the 2022 IEEE Signal Processing Magazine best paper award, the 2022 IEEE Signal Processing Magazine best column award, and the 2023 EMNLP outstanding paper award. 
	Dr. Dong Yu was the chair of the IEEE Speech and Language Processing Technical Committee between 2021-2022 and the technical co-chair of ICASSP-2021. 
\end{IEEEbiography}
\begin{IEEEbiography}[{\includegraphics[width=1in,height=1.15in,clip,keepaspectratio]{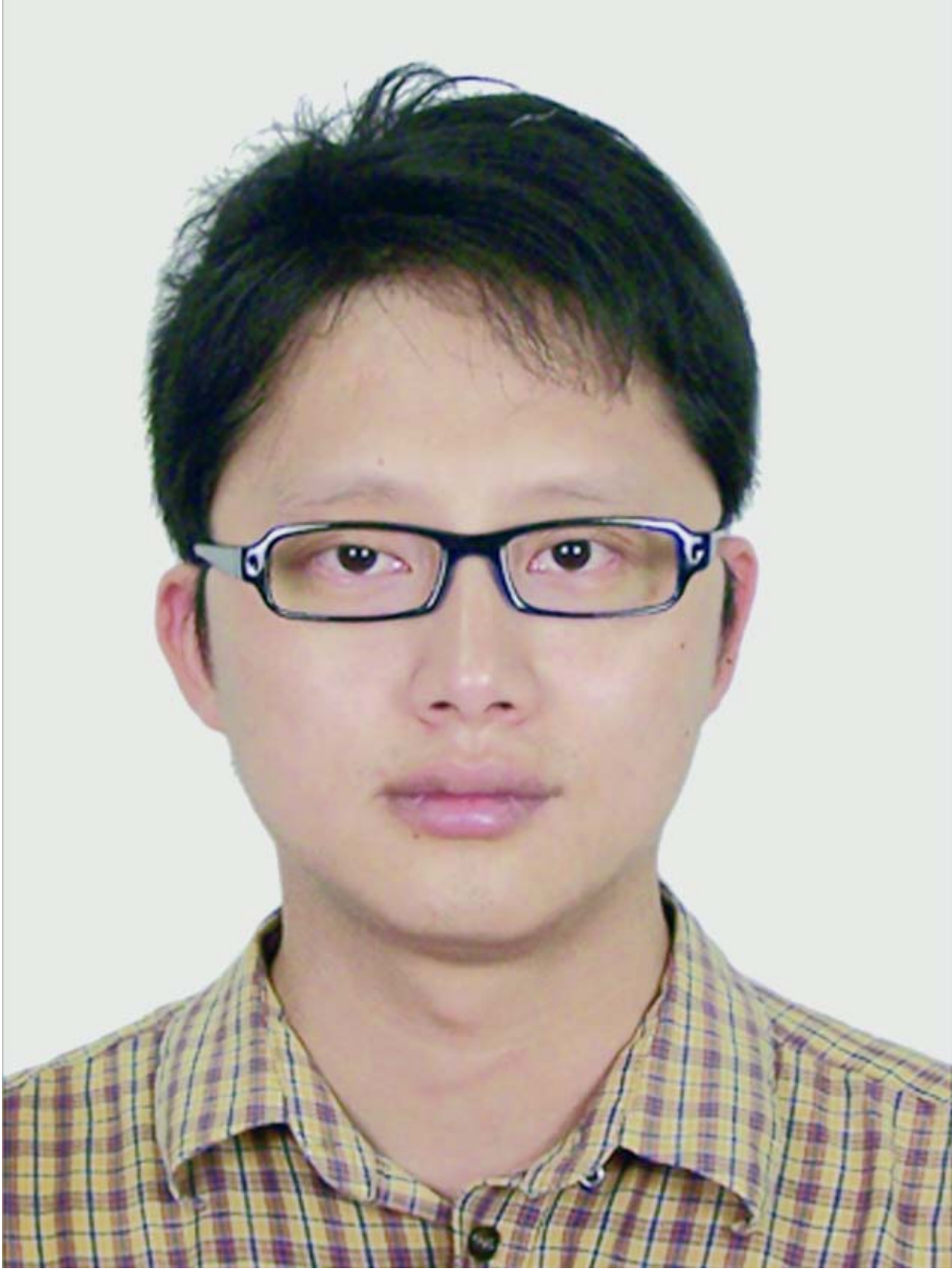}}]
	{Chengshi Zheng}
	(Senior Member, IEEE)
	received the B.S. degree in electronic engineering and information science from the University of Science and Technology of China, Hefei, China, in 2004, and the Ph.D. degree in acoustics from the Institute of Acoustics, Chinese Academy of Sciences, Beijing, China, in 2009. Since then, he has been with the Institute of Acoustics, Chinese Academy of Sciences, where he is currently a Full Professor with the Key Laboratory of Noise and Vibration Research. From 2014 to 2015, he was a Visiting Scientist with the Chair of Multimedia Communications and Signal Processing, University Erlangen-Nuremberg, Erlangen, Germany. His research interests include speech enhancement, deep learning, array signal processing, and signal processing for hearing aids. He was the recipient of the Outstanding Reviewer Awards at INTERSPEECH 2023.
\end{IEEEbiography}
\end{document}